\documentclass[tightenlines,floatfix,11pt]{article}

\usepackage{comment}
\usepackage{epsfig}
\usepackage{amsfonts}
\usepackage{amsmath}
\usepackage{amssymb}
\usepackage{dsfont}
\usepackage{color}
\usepackage[dvipsnames]{xcolor}
\usepackage[colorlinks, linkcolor=Maroon, citecolor=blue, filecolor=magenta, urlcolor=blue]{hyperref}
\usepackage{cite}
\usepackage[T1]{fontenc}

\renewcommand{\i}{\mathrm{i}}
\newcommand{\spacevec}[1]{\mathbf{#1}}
\newcommand{\pfrac}[2]{\left(\dfrac{#1}{#2}\right)}
\newcommand{\omegak}{\omega_{\spacevec k}}

\renewcommand{\d}[2]{%
  \ifthenelse{\equal{#1}{3}}%
    {\frac{d^3\spacevec{#2}}{(2\pi)^3}}%
    {\ifthenelse{\equal{#1}{4}}%
      {\frac{d^4{#2}}{(2\pi)^4}}%
      {d^{#1}#2}}%
}

\newcommand{\pext}{|\spacevec p|_{\rm ext}}

\newcommand{\boldtau}{\mbox{\boldmath $\tau$}}

\newcommand{\be}{\begin{eqnarray}}
\newcommand{\ee}{\end{eqnarray}}

\newcommand{\Order}{\mathcal{O}}

\newcommand{\MS}{$\overline{\text{MS}}$ }

\newcommand{\A}{\mathcal{A}}
\newcommand{\kk}{\mathbf{k}}

\newcommand{\dNLO}{d^{\rm NLO}}

\newcommand{\GeV}{\,\text{GeV}}
\newcommand{\MeV}{\,\text{MeV}}

\newcommand{\al}{\alpha}
\newcommand{\bt}{\beta}

\newcommand{\dslash}[1]{#1 \llap{/\kern-0.5pt}}
\newcommand{\Dslash}[1]{#1 \llap{/\kern+1.5pt}}
\newcommand{\DDslash}[1]{#1 \llap{/\kern+2.3pt}}
\newcommand{\dslashh}[1]{#1 \llap{/\kern+1pt}}

\newcommand{\bea}{\begin{align}}
\newcommand{\eea}{\end{align}}
\newcommand{\bma}{\begin{pmatrix}}
\newcommand{\ema}{\end{pmatrix}}

\newcommand{\NLDBD}{$0 \nu \beta \beta$}
\newcommand{\boldNLDBD}{$\boldsymbol{0\nu\beta\beta}$}

\numberwithin{equation}{section}

\allowdisplaybreaks[1]

\begin{document}
\begin{titlepage}

\begin{flushright}
INT-PUB-24-064
\end{flushright}

\vspace{1cm}

\begin{center}
{\LARGE  \bf  
Neutrinoless double $\boldsymbol{\beta}$ decay with  light  
\\[0.5cm]
sterile neutrinos:
the contact terms
\\[0.5cm]
}
\vspace{1cm}

{\large \bf  Vincenzo Cirigliano,$^a$ Wouter Dekens,$^a$
  Sebastián Urrutia Quiroga$^{a,b}$}
\vspace{0.5cm}

\vspace{0.25cm}

{\large 
$^a$ 
{\it Institute for Nuclear Theory, University of Washington, Seattle, WA 98195, USA}}

{\large 
$^b$ 
{\it Northwestern University, Department of Physics \& Astronomy, 2145 Sheridan Road, Evanston, IL 60208, USA}}

\end{center}

\begin{abstract}

We study neutrinoless double-beta decay in extensions of the Standard Model that include $n$ right-handed neutrino singlets, with masses 
$m_s$  
below the GeV scale.  
Generalizing recently developed matching methods, 
we determine the $m_s$ dependence of the short-range $nn \to pp$ couplings that appear to leading order 
in the chiral effective field theory description of neutrinoless double beta decay. 
We focus on two scenarios, corresponding to the minimal $\nu$SM and left-right symmetric models. 
We illustrate the impact of our new results in the case of the $\nu$SM, showing a significant 
impact on  the neutrinoless double-beta decay half-life  when  $m_s$ is in the 200-800 MeV range.

\end{abstract}

\vfill
\end{titlepage}

\tableofcontents

\section{Introduction}\label{sec:intro}

The Standard Model of particle physics (SM) in its original form \cite{Glashow:1959wxa,Salam:1959zz,Weinberg:1967tq} predicts massless neutrinos. Experimentally, however, neutrino masses have been convincingly established by oscillation experiments \cite{Superkamiokande1998,SNO:2001kpb,KamLAND:2002uet,ParticleDataGroup:2024cfk}, making the origin of neutrino masses one of the most important open questions in 
subatomic 
physics. Answering this question could shed light on other observations, such as the baryon asymmetry of the universe \cite{Fukugita:1986hr,Asaka:2005pn} and the presence of dark matter \cite{Dodelson:1993je,Asaka:2005pn}, that are left unexplained in the SM. 
From a theoretical perspective, in the absence of a symmetry principle, beyond-the-SM (BSM) scenarios that generate neutrino masses will generally lead to Majorana particles \cite{Weinberg:1979sa}. Neutrinoless double beta decay (\NLDBD), the observation of which would establish the neutrinos as Majorana particles, is, therefore, one of the most promising ways to gain insight 
into the nature of neutrino masses and address some of the most pressing open questions 
that find no answer in the SM. \\

In this paper, we focus on contributions to \NLDBD\ in minimal extensions of the Standard Model (SM) that involve, in addition to the SM fields,  a set of $n$ spin-1/2 singlets, the so-called sterile neutrinos.  
This scenario often called the $\nu$SM  \cite{Asaka:2005pn}, can explain nonzero neutrino masses and has been extensively studied in the literature as a possible explanation of the baryon asymmetry of the universe \cite{Asaka:2005pn,Shaposhnikov:2006nn,Shaposhnikov:2008pf,Canetti:2012kh,Canetti:2012vf,Drewes:2016jae,Drewes:2017zyw,Drewes:2021nqr} and dark matter \cite{Asaka:2005pn,Shaposhnikov:2008pf,Boyarsky:2018tvu}. The sterile neutrinos in the $\nu$SM are generally Majorana particles, which lead to a source of lepton-number violation and contributions to \NLDBD. The implications of light sterile neutrinos for  \NLDBD\  have been worked out in various frameworks \cite{Blennow:2010th,Mitra:2011qr,Li:2011ss,deGouvea:2011zz,Faessler:2014kka,Barea:2015zfa,Giunti:2015kza,Asaka:2005pn,Asaka:2011pb,Asaka:2013jfa,Asaka:2016zib,deVries:2022nyh,deVries:2024rfh}, here we will follow the effective-field-theory (EFT) 
approach set-up in Refs.~\cite{Dekens:2020ttz,Dekens:2023iyc,Dekens:2024hlz}. \\

The minimal scenario considered here extends the SM with $n$ $\nu_R$ singlets. 
Including operators up to dimension-4, this extends the Lagrangian by a kinetic term, a Majorana mass term, and their Yukawa interactions with the Higgs and lepton doublets.   
Expanding the Higgs field around its vacuum expectation value induces the Dirac mass term and, after 
diagonalizing the joint mass matrix for the active ($\nu_L$) {\it and} singlet ($\nu_R$) neutrinos, one ends up with $3+n$ Majorana mass eigenstates. This diagonalization can be achieved through a flavor rotation $N=U N^{(m)}$, where $N = (\nu_L,\nu_R^c)^{\sf T}$ collects the flavor eigenstates of the active and sterile neutrinos~\footnote{Here $\psi^c$ denotes the charge-conjugated field, $\psi^c = C \bar{\psi}^{\sf\, T}$, with $C$ the charge-conjugation matrix, which, in the Weyl representation, is given by $C=\i\gamma_2\gamma_0$.} and $N^{(m)}$ denote their mass eigenstates, while the $(3+n)\times (3+n)$ unitary matrix  $U_{\alpha j}$ is the generalization of the PMNS matrix that connects the neutrinos in the flavor and mass bases. The Majorana mass eigenstates can finally be written as $\nu_i = N^{(m)}_i+ \big(N^{(m)}_i\big)^c$ with $i = 1, \ldots, 3+n$~\cite{Dekens:2024hlz}. 
The relevant terms of the  effective Lagrangian at a scale $\mu   \gtrsim \Lambda_\chi$ are given by
\begin{align}
{\cal L}_{\rm eff}  &\supset  
 \frac{1}{2}\sum_{i=1}^{3+n} \bar \nu_i \left[\i  \protect{\slash\hspace{-0.5em}\partial}   - m_i\right]  \nu_i    
 \ - \  \left\{ \   2 \sqrt{2} G_F V_{ud} \    \bar{u}_L \gamma^\mu d_L \, \bar{e}_L \gamma_\mu \nu_{eL} 
\ +  {\rm h.c.}\right\}\,,
\label{eq:Seff0}
\end{align}
with $V_{ud}$ the $u-d$ component of the Cabibbo-Kobayashi-Maskawa (CKM) matrix~\cite{Cabibbo:1963yz,Kobayashi:1973fv}, $G_F$ the Fermi constant, and 
\begin{align}
\nu_{eL} &= P_L \, U_{ei} \nu_i\,,
\end{align}
where $P_L= (1 - \gamma_5)/2$.\\

Eq.~\eqref{eq:Seff0} induces $\Delta L= 2$ transitions by inserting two weak interactions. After contracting the neutrino fields in the two weak vertices, these LNV effects can be described by the following non-local effective action 
\begin{align}
 S_{\rm eff}^{\Delta L = 2} 
=\,& 
\frac{ 8  G_F^2 V_{ud}^2  
}{2!} 
\ \sum_{i= 1}^{ 3+n} U_{ei}^2\, m_i
\, \int d^4 x\,  d^4y   \ 
\bar{e}_L   (x)  \gamma^\mu  \gamma^\nu e_L^c (y)  
\nonumber \\
& \times 
 \int \frac{d^4k}{(2 \pi)^4}   \frac{e^{-\i k \cdot (x - y)}}{k^2 - m_i^2 + \i \epsilon} 
\  T \Big\{ J_\mu^L(x)   \ J_\nu^L(y)  \Big\}\,\,,
\label{eq:Seff1-v0}
\end{align}
where we have defined the left-handed current $J_\mu^L\equiv \bar u_L\gamma_\mu d_L$ 
and the charge-conjugated field $e_L^c = C \bar{e}_L^{\sf T}$.   
In the first three terms in the sum ($i=1,2,3$) one usually neglects $m_i^2$ in the denominators to recover the usual 
amplitude, proportional to $m_{\beta \beta} =  \sum_{i=1}^3 U_{ei}^2\, m_i$.  It should be noted, however, that in the $\nu$SM the sum over all neutrinos satisfies $ \sum_{i=1}^{3+n} U_{ei}^2\, m_i=0$. The $m_i^2$ in the denominators, therefore, cannot be neglected for the remaining $n$ terms in the sum, even when $m_i^2\ll k^2$, and we keep their mass non-zero throughout the analysis.\\

Although the representation in Eq.~\eqref{eq:Seff1-v0} is general, it is hard to evaluate starting from first principles. In order to take matrix elements between nuclear states, it is therefore useful to match this effective action to that of Chiral Perturbation 
Theory ($\chi$PT)~\cite{Weinberg:1978kz,Gasser:1983yg,Gasser:1984gg} and Chiral Effective Field Theory ($\chi$EFT)~\cite{Weinberg:1990rz,Weinberg:1991um}, 
involving pions and nucleons as degrees of freedom. This theory is based on an expansion in $Q/\Lambda_\chi$, where $Q$ corresponds to scales of the order of $m_\pi$ or the Fermi momentum, $k_F$, and $\Lambda_\chi\sim m_N\sim 1$ GeV is the breakdown scale of $\chi$PT. In addition, for nonzero neutrino masses, the $\chi$PT description also relies on an expansion in $m_s/\Lambda_\chi$. This additional expansion arises from the fact that positive powers of $m_s$ can arise from loop diagrams. Such diagrams are then no longer necessarily suppressed by factors of $Q/(4\pi F_\pi)\sim Q/\Lambda_\chi$, but another ratio, $m_s/(4\pi F_\pi)$, becomes possible. Whenever $m_s\sim \Lambda_\chi$, this expansion becomes unreliable, and for $m_s > \Lambda_\chi$, the sterile neutrino should be integrated out of the low-energy effective theory.

The chiral Lagrangian receives contributions that can be organized by the momentum and mass of the exchanged sterile neutrino:
\begin{itemize}
\item For heavy sterile neutrinos, $m_s\gtrsim \Lambda_\chi\simeq 1$ GeV, the neutrino can be integrated out at the quark level. The only relevant momentum region, with $k\gtrsim \Lambda_\chi$, in Eq.\ \eqref{eq:Seff1-v0} then results in effective dimension-nine operators, $O^{(9)}\sim (\bar ee^c)(\bar ud)^2$, at the quark level that can be matched onto $\chi$PT, leading to $\pi\pi$, $\pi N$, and $N\!N$ interactions.
\item 
For intermediate $k_F\lesssim m_s\lesssim \Lambda_\chi$ and light masses $m_s\lesssim k_F$, the sterile neutrinos have to be kept as explicit degrees of freedom within $\chi$PT. Here, several $\nu_R$ momentum regions can contribute:
\begin{enumerate}
\item The hard region with $k_0\sim |\spacevec k| \gtrsim \Lambda_\chi$ induces a contact interaction involving two nucleons, $\sim(\bar ee^c)(\bar pn)^2$.
\item The potential region with $k_0\ll|\spacevec k| \sim k_F\sim m_\pi$ induces the well-known long-range potential between nucleons, which takes the form of a Yukawa potential for massive neutrinos, $V(\spacevec q)\sim 1/(\spacevec q^2+m_s^2)$ in momentum space or $V(\spacevec r)\sim e^{-m_s r}/r$ in coordinate space.
\item Finally, neutrinos with ``ultrasoft'' momenta, $k_0\sim |\spacevec k|\sim k_F^2/m_N\sim {\cal Q}\sim 1$ MeV, where ${\cal Q}$ is a scale of the order of the $Q$ value of \NLDBD\ or excitation energies of the nuclei.
\end{enumerate}
\end{itemize}

In the mass region with $m_s\gtrsim \Lambda_\chi$ the resulting interactions in $\chi$PT are determined by hadronic matrix elements, some of which have been determined using lattice QCD computations \cite{Nicholson:2018mwc,Detmold:2022jwu}. Importantly, the $m_s$ dependence can be determined analytically in this mass region.
Here, we will focus on scenarios with lighter sterile neutrinos $m_s\lesssim \Lambda_\chi$, in which case, non-perturbative hadronic and/or nuclear information is needed to determine the $m_s$ dependence. In particular, the contributions of the momentum regions $(2)$ and $(3)$ require input from nuclear structure, either from the matrix element of the LNV potential in region (2) or through information about intermediate states in region (3). These effects are expected to be dominant for $m_s\lesssim k_F$ and have been discussed in detail in Refs.\ \cite{Dekens:2023iyc,Dekens:2024hlz}.\\ 

In this paper, we focus instead on region $(1)$, whose contributions are expected to be most significant when $\Lambda_\chi\gtrsim m_s\gtrsim k_F$. As this region is generated at scales of order $\Lambda_\chi$, the $m_s$ dependence is determined by hadronic physics. 
In the EFT, these effects are parametrized by the mentioned contact interaction, which is needed to renormalize the contributions to the $nn \to pp$ transition from potential-neutrino exchange contributions of region (2). This interaction is completely analogous to the contact term needed in the case of active neutrino exchange with $m_s=0$~\cite{Cirigliano:2019vdj}.\\

To  leading order in $Q/\Lambda_\chi$, 
the hard neutrino modes of region (1) discussed above induce local operators in the low-energy EFT
\begin{align}
\mathcal L_{\Delta L =2}^{N\!N} &= \left( 2 \,  G_F^2 V_{ud}^2  \right) 
\,   \sum_{i=1}^{3+n} \, U_{ei}^2  \, m_i    \, {\cal C}_1 (m_i)  \times
  \bar{e}_L e_L^c    \,   \bar N \mathcal \tau^+ N \, \bar N \tau^+ N\,.   
  \label{eq:C1v1}
\end{align}

In the limit $m_i \to 0$, 
the LNV coupling  ${\cal C}_1 (m_i) \equiv g_\nu^{N\!N} (m_i)$ 
coincides with the coupling  introduced in Ref.~ \cite{Cirigliano:2019vdj} 
and estimated in the Cottingham approach in Refs.~\cite{Cirigliano:2020dmx,Cirigliano:2021qko}. In $\chi$PT LECs can only exhibit polynomial dependence on $m_s$, while any non-analytic 
dependence of  amplitudes 
can only result from loop diagrams. Strictly speaking therefore, the coupling ${\cal C}_1$ is a linear combination of $m_s$-independent LECs, \emph{i.e.,}\ ${\cal C}_1(m_s) =\sum _{n=0}{ c}^{(n)}_1 m_s^n$. Renormalization arguments show that the $n=0$ coefficient is needed at leading order, $c_1^{(0)}={\cal O}(F_\pi^{-2})$, while no such arguments hold for the $m_s$-dependent terms. These corrections are expected to scale as $c_1^{(n)}={\cal O}(F_\pi^{-2}\Lambda_\chi^{-n})$ from a naive-dimensional-analysis (NDA) \cite{Manohar:1983md,Weinberg:1989dx} estimate.\\

The main objective of this work will be to determine the $m_s$ dependence of the coupling ${\cal C}_1 (m_s)$ by matching the effective $\chi$PT description of $nn\to pp$ to a modeled version of Eq.\ \eqref{eq:Seff1-v0}. As this low-energy constant (LEC), ${\cal C}_1$, captures non-perturbative hadronic physics, it could, in principle, be determined from lattice QCD calculations. Here, however, 
we 
extend the methodology of Refs.~\cite{Cirigliano:2020dmx,Cirigliano:2021qko}, to which we refer the reader for details. 
While we will not repeat all the technical steps common with Refs.~\cite{Cirigliano:2020dmx,Cirigliano:2021qko}, 
we will discuss  the significant changes and modifications to the matching procedure 
that emerge in the presence of new light states in the theory. \\

In Section~\ref{sec:ALL_ms}   we will determine  the dependence of  ${\cal C}_1 (m_i)$  on the sterile neutrino mass  
in the regime $m_s <  m_K \sim 3 m_\pi$,  such that   $m_s/\Lambda_\chi$ can be  treated an  expansion parameter. 
In Section~\ref{sec:AVV_ms}, we will then generalize the analysis to non-minimal new physics scenarios in which the neutrinos 
couple to the right-handed quark current, corresponding to  
changing one of the currents in  \eqref{eq:Seff1-v0}
from $J_\mu^L\equiv \bar u_L\gamma_\mu d_L$ to  $J_\mu^R\equiv \bar u_R\gamma_\mu d_R$.
In Section~\ref{sec:impact}, we will return to the minimal scenario and study the impact of the 
newly determined ${\cal C}_1 (m_i)$ on neutrinoless double beta decay rates, showing a significant effect in the 
region around $m_i \sim 100$~MeV. 
Finally, we present our conclusions in Section~\ref{sec:conclusion}. 
Some more technical details are relegated to the Appendices.

\section{The minimal scenario: Left-Left amplitude}
\label{sec:ALL_ms}

To determine  ${\cal C}_1 (m_i)$, we carry out a matching calculation at the level of the amplitude for the process 
$n ({\bf p}) n ( - {\bf p})   \to p ({\bf p}^\prime)  p (- {\bf p}^\prime)  e^- ({\bf p}_e)  e^- (- {\bf p}_e)$, 
\begin{align}
\langle e_1^- e_2^- pp |   S_{\rm eff}^{\Delta L = 2}  | nn \rangle 
&=  (2 \pi)^4 \, \delta^{(4)}  (p_f - p_i)  
\Big( 4  G_F^2 V_{ud}^2 
\   \bar{u}_L   (p_1)  u_L^c (p_2)  \Big)  
\times 
\sum_{i=1}^{3+n} U_{ei}^2 \, m_i  \, 
 {\cal A}_\nu(m_i)\,,\notag   \\
  {\cal A}_\nu(m_i)  & =
2\,  g^{\mu \nu}  \, 
 \int \frac{d^4k}{(2 \pi)^4} \, 
\int d^4 r  \,
e^{i k \cdot r}
\, \frac{   
\langle pp | \,  
T  \Big\{  J_\mu^L (+r/2)   \ J_\nu^L(- r/2)  \Big\}  
| nn \rangle 
 }{ k^2 - m_i^2 + \i \epsilon}\, . 
  \label{eq:M}
\end{align}

Note that since the contributions of the $n$ additional neutrinos are additive, we will  focus on  a single sterile  neutrino 
and will denote its mass by $m_s$. Moreover,  it is convenient to carry out the analysis in terms of dimensionless hadronic 
amplitudes defined via 
\be
\bar {\cal A}_\nu (m_s)
 \equiv  \left( \frac{4 \pi}{m_N} \right)^2 \  {\cal A}_\nu (m_s) \,~. 
\label{eq:match4}
\ee

The hadronic amplitude ${\cal A}_\nu (m_s)$ in Eq.~\eqref{eq:M} receives contributions from a wide range of neutrino virtualities $k^2$. To calculate this amplitude, we have decomposed $d^4 k = d k^0 \,d^3\spacevec{k}$ to first perform the $k^0$ integral (quite natural given 
the presence of non-relativistic nucleons) and then carried out the angular integrations in $d^3\spacevec{k}$ to reduce the amplitude to an integral over $d |\spacevec{k}|$. 
We then split  the integral into a low- plus intermediate-momentum region and a high-momentum region 
\begin{align}
\bar {\cal A}_\nu^{\rm full}(m_s) &= 
 \int_0^\infty \,  d |\spacevec{k}|  \ a^{\rm full}  (|\spacevec{k}|,m_s) = 
\bar  {\cal A}^<(m_s) + \bar { \cal A}^>(m_s)\,,  
\notag
\\
\bar {\cal A}^<(m_s)  &=    \int_0^{\Lambda} \,  d |\spacevec{k}|  \ a_< (|\spacevec{k}|,m_s)\,,\notag
\\
\bar { \cal A}^>(m_s)  &=  \int_{\Lambda}^\infty \,  d |\spacevec{k}|  \ a_> (|\spacevec{k}|,m_s)\,, 
\label{eq:Msplit}
\end{align}
separated by an arbitrary scale $\Lambda$,  at which  the asymptotic behavior  for the current--current correlator, controlled by the 
operator product expansion (OPE) sets in.  The basic idea behind the Cottingham approach is that model-independent representations of the integrands in Eq.~\eqref{eq:Msplit} can be constructed in the low-energy, $|\spacevec{k}|\ll\Lambda_\chi$, region (via pionless and chiral EFT) and in the hard, $|\spacevec{k}|\gg\Lambda_\chi$, region (via the OPE), whereas a model for the full amplitude can be constructed by interpolating between these two regions. Once a representation for $\bar {\cal A}_\nu^{\rm full}$ is obtained, we estimate the LEC  ${\cal C}_1 (m_i)$ appearing in $ \bar {\cal A}_\nu^{\chi {\rm EFT}}$ by enforcing the matching condition 
\begin{align}
\bar {\cal A}_\nu^{\chi {\rm EFT}}(m_s) =  \bar {\cal A}^<(m_s) + \bar  { \cal A}^>(m_s) \,.
\label{eq:match0}
\end{align}

This approach, while partly model-dependent due to the need to interpolate in the nonperturbative regime 
of Quantum ChromoDynamics (QCD), has been validated at the $\sim 30\%$ level 
in the case of the nucleon-nucleon electromagnetic contact terms~\cite{Cirigliano:2020dmx,Cirigliano:2021qko}.

\begin{figure}[t]
\centering
\includegraphics[width=\linewidth,clip]{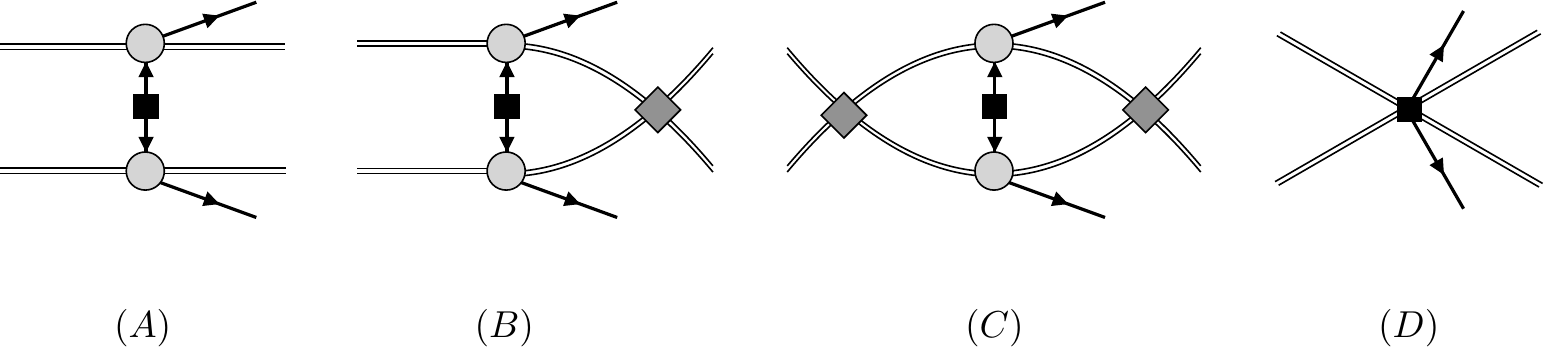}
\caption{
Topologies relevant for the $nn \to pp e^-e^-$ amplitude to leading order in the chiral expansion. 
The 
double 
solid lines denote nucleons, while the thin solid oriented lines denote  
leptons (the internal neutrino and the external electrons).  
The black squares denote the lepton-number violating vertices: Majorana mass insertions in topologies 
$A$, $B$, and $C$, and the contact interaction ${\cal C}_1 (m_s)$ in topology $D$. 
The gray circles represent the leading order $V-A$ weak current insertions. The gray diamond denotes the leading-order contact interaction of two nucleons in the $^1 S_0$ channel. 
The same topologies contribute to the amplitude in the underlying theory (weak interactions plus QCD) with different meanings for the diamonds and circles. 
}
\label{fig:diagrams}
\end{figure}

\subsection[Matching and extraction of ${\cal C}_1 (m_s)$]{Matching and extraction of $\boldsymbol{{\cal C}_1 (m_s)}$}
\label{sec:matching}

For any value of $m_s \leq \Order(\Lambda_\chi)$, 
the  chiral EFT $nn \to ppe^-e^-$  amplitude  can be decomposed as 
\be\label{eq:AchiPT}
\bar \A_\nu^\text{EFT}=\bar \A_A+ \bar \A_B+ \bar \A_C+ \bar \A_D,
\ee
with each of  the four terms corresponding to the topologies 
shown in  Fig.~\ref{fig:diagrams}. 
To leading order 
in chiral EFT, the gray circles represent insertions of the leading-order $V-A$ weak current 
and the gray diamond denotes the leading-order strong contact interaction 
of two nucleon in the $^1 S_0$ channel, proportional to the effective coupling $C(\mu_\chi)$.
Even after renormalizing the strong interactions $\bar \A_C$ remains divergent, which requires the 
leading order (LO) contact term $\bar \A_D \propto {\cal C}_1$~\cite{Cirigliano:2018hja}, 
with ${\cal C}_1$ defined in Eq.~\eqref{eq:C1v1}.
In the full theory, the amplitude can be organized according to  the same topologies appearing in Fig.~\ref{fig:diagrams}, 
by replacing the fixed order EFT vertices with 
the appropriate form factors that 
capture the momentum dependence of the  $N\!N$ intermediate-state contributions.
Additional iterations of the $N\!N$ strong Yukawa interactions (pion exchange) and short-range interactions (diamonds) 
are not shown as they are irrelevant for the matching analysis.  \\

As discussed in detail in Refs.~{\cite{Cirigliano:2020dmx,Cirigliano:2021qko}, 
for matching purposes, only topologies ($C$) and ($D$) are relevant, as the others coincide in chiral EFT and underlying theory up to higher-order effects. 
This continues to hold in the presence of massive neutrinos as long as 
one neglects terms of order $m_s/\Lambda_\chi$. 
In particular, only the ultraviolet singular part of topology ($C$), 
involving noninteracting two-nucleon propagators (\emph{i.e.,} without pion exchanges) 
enters the matching condition. 
Defining the dimensionless coupling $\tilde{\cal C}_1$ via
\begin{align}
{\cal C}_1 (\mu_\chi,m_s)  &=   \left( \frac{m_N}{4 \pi} C (\mu_\chi) \right)^2  \ \tilde{\cal C}_1 (\mu_\chi,m_s)\,,
\label{eq:match3}
\end{align}
where $C(\mu)$ is the leading non-derivative $N\!N$ coupling in the $^1 S_0$ channel in the convention of Refs.~{\cite{Cirigliano:2020dmx,Cirigliano:2021qko}, the matching condition reads
\begin{align}
\label{matching}
\bar \A_C^{<, \text{sing}}(m_s) + \bar \A_C^{>}(m_s) &=
\bar \A_C^{\text{sing}}(\mu_\chi,m_s)+2\tilde {\cal C}_1(\mu_\chi,m_s)\,.
\end{align}

The left-hand side of Eq.~\eqref{matching} denotes the full amplitude, separated into momentum  regions 
\begin{align}
\bar{\cal A}^{<,{\rm sing}}_{C}(m_s) &=  \int_0^{\Lambda} \  d |\spacevec{k}| \ a_< ( | \spacevec{k} |,m_s) \,,\notag
\\
\bar{\cal A}^{>}_{C}(m_s) &=   \int_{\Lambda}^\infty   \  d |\spacevec{k}| \ a_> ( | \spacevec{k} |,m_s) \,. 
\label{eq:integrals}
\end{align}

On the other hand, the right-hand side 
of Eq.~\eqref{matching} 
gives the amplitude in chiral EFT, including the contact term $\tilde {\cal C}_1$ at the \MS scale $\mu_\chi$ (we use here the notation of Ref.~\cite{Cirigliano:2019vdj}) 
and the UV singular loop amplitude~\footnote{There is an implicit separation into real/imaginary 
parts of Eq.~\eqref{matching}, and the matching condition considers only the real part of $\mathcal A_C^X$.}. \\

In generalizing the matching analysis of Refs.~\cite{Cirigliano:2020dmx,Cirigliano:2021qko}  
to include massive neutrinos (with $m_s \neq 0$),  
several subtleties arise, related to the  IR behavior of the full and EFT amplitudes.  
The principles of EFT require that 
the effective theory should have the same IR behavior as the full theory. 
In turn, 
this implies that 
(i) $\tilde{\cal C}_1 (m_s)$  does not depend on the  momenta of the incoming $nn$ or outgoing $pp$ pairs 
(the effective couplings do not depend on the external states used in the matching condition);  
(ii)  
$\tilde{\cal C}_1$  depends on $m_s$  only through a polynomial in $m_s^2$~\cite{Dekens:2024hlz}, 
\begin{align}
{\cal C}_1(m_s) &= {\cal C}_1^{(0)} + {\cal C}_1^{(2)}\,m_s^2 + \Order(m_s^4)\,.
\label{eq:gvNN(ms)}
\end{align}

This can be understood as follows: for the mass ranges we consider in this work, sterile neutrinos are active degrees of freedom in the low-energy EFT, and hence, the non-analytic behavior in $m_s$ (\emph{e.g.,} $\log m_s$) arises from loops in the EFT. Therefore, one would expect that any non-analytic behavior appearing on the full theory side of Eq.\ \eqref{matching} should be canceled by loop diagrams on the chiral EFT side.
In what follows, we discuss both sides of the matching equation in greater detail and report the final results for $\tilde {\cal C}_1 (\mu_\chi,m_s)$ in Eq.~\eqref{eq:mc4}.

\subsubsection{Full amplitude}
\label{sect:fullLL}

A generalization of the analysis of Refs.~\cite{Cirigliano:2020dmx,Cirigliano:2021qko} 
to include non-zero neutrino mass leads to a 
determination of $a_< ( |\spacevec{k} |,m_s)$ and $a_> ( |\spacevec{k} |,m_s)$, which we now discuss. \\

\noindent  {\bf High momentum region:}  For $ | \spacevec{k} |>  \Lambda $ using the OPE we arrive at the expression 
\begin{equation}
a_> ( | \spacevec{k} |,m_s)    = 
 \frac{4 \alpha_s}{\pi} \ \bar{g}_1^{N\!N} \ F_\pi^2 \ \frac{2|\spacevec{k}| + \omegak}{\omegak|\spacevec{k}|(|\spacevec{k}| + \omegak)^2} \,, 
 \label{eq:aLLp}
\end{equation}
where $\omegak = \sqrt{\spacevec{k}^2 + m_s^2}$ and  $ \bar{g}_1^{N\!N}$ 
is proportional to  the  $nn \to pp$  matrix element of the local operator $\bar u_L \gamma^\mu d_L \bar u_L \gamma_\mu d_L$.  
Since this matrix element is currently unknown and naive dimensional analysis suggests 
$ \bar{g}_1^{N\!N} \sim \Order(1)$, we take the range $ \bar{g}_1^{N\!N} \in [-10,+10]$. 
As in the massless neutrino case~\cite{Cirigliano:2020dmx,Cirigliano:2021qko},
even using this conservative estimate, the high momentum region gives a 
very small 
contribution 
to $\tilde {\cal C}_1 (\mu_\chi, m_s)$. \\

\noindent  {\bf Small and intermediate momentum region:} 
For the $ | \spacevec{k} | <   \Lambda $ region,  following the steps in 
Refs.~\cite{Cirigliano:2020dmx,Cirigliano:2021qko} we obtain
\begin{equation}
a_< ( | \spacevec{k} |,m_s)   = -8\, g_{\rm full}({\bf k}^2)
 \,     \frac{\spacevec{k}^2}{\spacevec{k}^2 + m_s^2}\ \operatorname{Re}\,\mathcal{I}_C^<(|\spacevec k|)\,,
\label{eq:aLLm}
\end{equation}
where we have defined
\begin{eqnarray}
g_{\rm full}({\bf k}^2) &\equiv& g_V^2  (\spacevec{k}^2) + 2 g_A^2( \spacevec{k}^2 ) +   \frac{\spacevec{k}^2 g_M^2 (\spacevec{k}^2)}{2 m_N^2}\,, \label{eq:gFull}
\end{eqnarray}
and expressions for the vector, axial, and magnetic form factors can be found in Ref.~\cite{Cirigliano:2021qko}.  
The quantity  ${\cal I}_C^<  (|\spacevec{k}|)$ is related to the two-nucleon bubble in topology ($C$) of Fig.~\ref{fig:diagrams}, 
\begin{align}
{\cal I}_C^<  (|\spacevec{k}|)   
&= 
 \int \frac{d^3 \spacevec{q}}{(2 \pi)^3}  \ 
 f_S   (\spacevec{p}^\prime, \spacevec{q} + \spacevec{k}) \ 
  \frac{1}{ {\bf p}^{\prime 2} - ( {\bf q}  + {\bf k})^2 + \i \epsilon} 
 \  
\frac{1}{ {\bf p}^{2} - {\bf q}^2 + \i \epsilon}
 \ 
 f_S  (\spacevec{q}, \spacevec{p} )\,,
\label{eq:ACmsing}
\end{align}
and $f_S (\spacevec{q}, \spacevec{p} )$ encodes the momentum-dependence of the short-range 
 (not mediated by pions) 
$^1S_0$ scattering amplitude, namely  
\begin{eqnarray}
f_S (\spacevec{q}, \spacevec{p} ) 
 = 
 \frac{T_S(\spacevec{q}, \spacevec{p})} {T_S (\spacevec{p}, \spacevec{p})}    \bigg \vert_{\alpha_\pi =0} 
 \,,
\label{eq:fsbarnopi} 
\end{eqnarray}
where 
$T_S(\spacevec{q}, \spacevec{p}) = \langle \spacevec{q}|\hat T_S(E)|\spacevec{p}\rangle$
is the  half-off-shell (HOS) matrix element~\cite{Srivastava:1975eg,Cirigliano:2021qko} 
of the scattering operator $\hat T_S$ induced by the short-range strong interaction.  
\\

At low-momentum, one can use chiral EFT to compute $f_S (\spacevec{q}, \spacevec{p} ) = 1  + ...$, 
while, at intermediate momenta, models for the strong $N\!N$ potential in the $^1 S_0$ channel~\cite{Kaplan:1999qa, Reid:1968sq, Wiringa:1994wb}
were used to carry out the calculation~\cite{Cirigliano:2021qko}. 
The expression to LO and NLO in  chiral EFT  at low-momentum are:
\begin{align}
\operatorname{Re}\,\mathcal{I}_C(|\spacevec k|) &\equiv \operatorname{Re}\,\mathcal{I}_C^<(|\spacevec k|) \Big \vert_{
f_S (\spacevec{q}, \spacevec{p} ) =1} 
=   \frac{1}{8|\spacevec k|}\,\theta(|\spacevec k| - |\spacevec p|_{\rm ext}) 
\\
\operatorname{Re}\,\mathcal{I}_C^<(|\spacevec k|)\big \vert_{\rm NLO}  &=
\frac{1}{8|\spacevec k|}\Big[\theta(|\spacevec k| - |\spacevec p|_{\rm ext}) - \dNLO\,|\spacevec k|\Big]\,,\label{eq:IC_lowmomentum}
\end{align}
with~\cite{Cirigliano:2021qko}
\begin{align}
\label{eq:dNLO}
\dNLO &\equiv \frac{8C_2}{m_NC^2}\quad,\qquad C_2/C^2\simeq0.38\text{\cite{Kaplan:1996xu}}\,.
\end{align}

Eq.~\eqref{eq:aLLm} reproduces the result in Eqs.\ (5.4)--(5.6) in Ref.~\cite{Cirigliano:2021qko} in the limit of $m_s\to0$ after trading 
$ \operatorname{Re}\,\mathcal{I}_C^<(|\spacevec k|) \to \operatorname{Re}\,\mathcal{I}_C (|\spacevec k|)\ r(|\spacevec{k}|)$, with 
\begin{align}
r( |\spacevec{k}|) &\equiv \frac{\operatorname{Re} \, {\cal I}_C^< (| \spacevec{k}|)}{  \operatorname{Re} \, {\cal I}_C (| \spacevec{k}|) }\,.
\label{eq:r0}
\end{align}

However, it turns out that  the replacement $ \operatorname{Re}\,\mathcal{I}_C^<(|\spacevec k|) \to\operatorname{Re}\,\mathcal{I}_C (|\spacevec k|)\ r(|\spacevec{k}|)$  
obscures the analysis of the small $m_s$ and small $|\spacevec{p}|_{\rm ext}$ regions of the amplitude, 
so we will use Eq.~\eqref{eq:aLLm} in what follows. \\

The coupling $\tilde{\cal C}_1$ extracted from the 
matching relation in Eq.~\eqref{matching}  should be independent of the external momenta $\pext$ 
and should have a polynomial dependence on $m_s^2$. 
To verify this statement, we need to isolate the $\pext$  dependence of 
$\operatorname{Re}\,{\cal I}_C^<$ 
in $\bar{\cal A}_C^{<,{\rm sing}}$ (full theory amplitude) 
that eventually will cancel out the $\pext$-dependence in $\bar{\cal A}_C^{\rm sing}$ (EFT amplitude).
To identify the $\pext$ dependence of $\operatorname{Re}\,{\cal I}_C^<$ analytically, 
it is convenient to introduce an auxiliary energy scale $\lambda$ such that 
$|\spacevec p|_{\rm ext} \ll \lambda\lesssim m_\pi$, and 
rewrite the first line in Eq.~\eqref{eq:integrals} as
\begin{align}
\bar{\cal A}^{<,{\rm sing}}_{C}(m_s) =& \int_0^{\lambda} \  d |\spacevec{k}| \ a_< ( | \spacevec{k} |,m_s) + \int_{\lambda}^{\Lambda} \  d |\spacevec{k}| \ a_< ( | \spacevec{k} |,m_s)\,.
\label{eq:decomposition_lambda}
\end{align}

The analysis of the two regions proceeds as follows:
\begin{itemize}
\item \textbf{Region $\boldsymbol{ |\kk| < \lambda}$:} 
Since $\lambda \lesssim  m_\pi  \ll\Lambda_V,\,\Lambda_A\sim\Order(1\,{\rm GeV})$, we can safely neglect the dipole effects in the form factors, leading to the approximation $g_{\rm full}(\spacevec k^2)\approx g_{\rm full}(0)=(1+2g_A^2)$.
Subsequently using the NLO expansion in Eq.~\eqref{eq:IC_lowmomentum} for  $\operatorname{Re}\,\mathcal{I}_C^<(|\spacevec k|)$, 
which is appropriate for $|\spacevec{k}| \lesssim \lambda$, one arrives at: 
\begin{align}
\int_0^{\lambda} \  d |\spacevec{k}| \ a_< ( | \spacevec{k} |,m_s)  &= 
\frac{(1+2g_A^2)}{2}\Bigg[2\dNLO\,\lambda - 2\dNLO\,m_s\tan^{-1}\pfrac{\lambda}{m_s} 
+\ \log\left( \frac{m_s^2+|\spacevec p|_{\rm ext}^2}{m_s^2+\lambda^2} \right)
\Bigg]\,.
\label{eq:A<1}
\end{align}

\item \textbf{Region $\boldsymbol{\lambda\leq |\spacevec k|\leq\Lambda}$:}
Since in this region the condition in the $\theta$ function in Eq.\ \eqref{eq:IC_lowmomentum} is always satisfied, we have $\operatorname{Re}\,{\cal I}_C(|\kk|)>0$ and we can safely re-introduce the ratio
\begin{align}
r( |\spacevec{k}|) &= \frac{\operatorname{Re} \, {\cal I}_C^< (| \spacevec{k}|)}{  \operatorname{Re} \, {\cal I}_C (| \spacevec{k}|) }\,, 
\notag
\end{align}

such that $a_<(|\spacevec k|)$ becomes:
\begin{align}
a_<(|\spacevec k|,m_s) =& -g_{\rm full}({\bf k}^2)\, \frac{|\spacevec{k}|\ r(|\spacevec{k}|)}{\spacevec{k}^2 + m_s^2}\ \theta(|{\bf k}| - |{\bf p}|_{\rm ext}) \,,\label{eq:a<_r0}
\end{align}

and the second term in the decomposition in Eq.~\eqref{eq:decomposition_lambda} is just
\begin{align}
\int_{\lambda}^{\Lambda} \  d |\spacevec{k}| \ a_< ( | \spacevec{k} |,m_s) &= -\int_{\lambda}^{\Lambda} \  d |\spacevec{k}| \ g_{\rm full}({\bf k}^2)\, \frac{|\spacevec{k}|\ r(|\spacevec{k}|)}{\spacevec{k}^2 + m_s^2}\,.
\label{eq:kgtl}
\end{align}
\end{itemize}

\subsubsection{EFT amplitude} 
\label{sec:ALL-subsec:IRpext-subsubsec:IR}

The relevant LO amplitude  in chiral EFT takes the form 
\begin{align}
\bar{\cal A}^{\rm sing, LO}_{C}  (\mu_\chi,m_s) &= -\frac{(1+2g_A^2)}{2}\left[1+\log\pfrac{\mu_\chi^2}{m_s^2+\pext^2}\right] \,.
\label{eq:chiEFT}
\end{align}

Note that Eqs.~\eqref{eq:chiEFT} and \eqref{eq:A<1} display the same  dependence on $| \spacevec{p}|_{\rm ext}$, 
and hence  $|\spacevec{p}|_{\rm ext}$ drops out in the matching condition for $\tilde {\cal C}_1$, a desirable feature. 
However, the linear term $\sim\dNLO\,\pi\,m_s$  in  \eqref{eq:A<1} has no analog in the EFT expression. 
This apparent mismatch comes from the fact that, on the full-theory side, the small-$|\spacevec k|$ regime generates 
an $m_s$ dependence proportional to $\dNLO$ (see Eq.~\eqref{eq:IC_lowmomentum} 
and subsequent discussion) that is not captured by the leading order chiral EFT description 
in Eq.~\eqref{eq:chiEFT}. As shown in  Appendix~\ref{sec:appNLO},  this dependence on $m_s$  arises on the EFT side at NLO:
\begin{align}
\bar{\cal A}^{\rm sing,NLO}_{C}  (\mu_\chi,m_s) &= -\frac{(1+2g_A^2)}{2}\Bigg[1+\log\pfrac{\mu_\chi^2}{m_s^2+|\spacevec p|^2_{\rm ext}} + \dNLO\,\pi\,m_s\Bigg]\,.
\label{eq:AchiNLO}
\end{align}

As we show below, including this NLO dependence on $m_s$ on the EFT side of the matching relation in Eq.~\eqref{matching} leads 
to an expression for $\tilde {\cal C}_1 (m_s)$ consistent with the expected form in Eq.~\eqref{eq:gvNN(ms)}.

\subsubsection[Expression for $\tilde {\cal C}_1 (m_s)$ and its small $m_s$ expansion]{Expression for $\boldsymbol{\tilde {\cal C}_1 (m_s)}$ and its small $\boldsymbol{m_s}$ expansion}

Using Eqs.~\eqref{eq:A<1}, \eqref{eq:kgtl}, and \eqref{eq:AchiNLO} in the matching condition in Eq.~\eqref{matching} we arrive at 
our main result:  
\begin{align}
\tilde{\cal C}_1  (\mu_\chi,m_s) 
=& \frac{1}{2}\Bigg\{\frac{(1+2g_A^2)}{2}\Bigg[  1  + \log \pfrac{\mu_\chi^2}{m_s^2+\lambda^2}
+\dNLO\,\pi\,m_s  \left( 1 -  \frac{2}{\pi} \tan^{-1}\pfrac{\lambda}{m_s} \right)
+ 2\dNLO\,\lambda \Bigg]
 \notag\\
&
- \int_{\lambda}^{\Lambda} \  d |\spacevec{k}| \ g_{\rm full}({\bf k}^2)\, \frac{|\spacevec{k}|\ r(|\spacevec{k}|)}{\spacevec{k}^2 + m_s^2}\notag\\
&
+  \frac{4 \alpha_s}{\pi} \, \bar{g}_1^{N\!N} \, F_\pi^2 
\int_{\Lambda}^\infty  d |\spacevec{k}| \, 
 \ \frac{2|\spacevec{k}| + \omegak}{\omegak|\spacevec{k}|(|\spacevec{k}| + \omegak)^2} \,, 
\Bigg\}\,,
\label{eq:mc4}
\end{align}
where the first, second, and third lines encode the effect from low ($|\spacevec{k}|< \lambda$), intermediate ($\lambda < |\spacevec{k}|< \Lambda$), and high ($\Lambda < |\spacevec{k}|$) neutrino momentum. 
We note the following features of this result: 
\begin{itemize}
\item As discussed below, 
$\tilde{\cal C}_1  (\mu_\chi,m_s) $ 
 is quite stable with respect to the choice of the arbitrary scales $\lambda$ and $\Lambda$, 
taken in the ranges  $|\spacevec{p}|_{\rm ext} < \lambda \lesssim m_\pi$  and $\Lambda \sim \Order({\rm GeV})$. 
\item $\tilde{\cal C}_1(\mu_\chi,m_s)$ does not have any logarithmic dependence on $|\spacevec{p}|_{\rm ext}$~\footnote{There 
is a very mild dependence  on $|\spacevec{p}|_{\rm ext}$,  arising from  $r(|\spacevec k|)$~\cite{Cirigliano:2021qko}, due to the condition $|\spacevec k|>|\spacevec p|_{\rm ext}$.\label{fnote:p-dep_r}}.
\item By expanding $\tilde{\cal C}_1  (\mu_\chi,m_s)$  in the regime $m_s \ll \lambda$ one can analytically recover the polynomial behavior in $m_s$. 
On the other hand, in the regime  $ \lambda \lesssim m_s$, the logarithmic dependence on $m_s$ cancels between the first and second lines of Eq.~\eqref{eq:mc4}.
\end{itemize}

Since the small $m_s$ behavior is of phenomenological interest~\cite{Dekens:2024hlz}, 
we provide the explicit expansion of  $\tilde{\cal C}_1$ in the IR limit of $m_s\ll\lambda<\Lambda$:
\begin{equation}
\tilde{\cal C}_1(\mu_\chi,m_s) =
\sum_{n=0} \tilde{\cal C}_1^{(2n)}m_s^{2n}\,,
\label{eq:taylorexp_C1_NLO}
\end{equation}
where the first few coefficients are given by
\begin{align}
\tilde{\cal C}_1^{(0)} &\equiv \frac{1}{2}\Bigg[\frac{(1+2g_A^2)}{2}\Bigg\{2\dNLO\,\lambda + 2\log\pfrac{\mu_\chi}{\lambda} + 1\Bigg\} - \alpha_0^< + \alpha_0^>\Bigg]\,,\notag\\
\tilde{\cal C}_1^{(2)} &\equiv \frac{1}{2}\Bigg[\frac{(1+2g_A^2)}{2}\Bigg\{\frac{2\dNLO}{\lambda} - \frac{1}{\lambda^2}\Bigg\} - \alpha_2^< + \alpha_2^>\Bigg]\,,
\label{eq:def_C10-C12}
\end{align}
and 
\begin{align}
\alpha_0^< &\equiv \int_{\lambda}^{\Lambda} \  d |\spacevec{k}| \ g_{\rm full}(\spacevec k^2)\,\frac{r(|\spacevec k|)}{|\spacevec k|}\,,\notag\\
\alpha_2^< &\equiv -\int_{\lambda}^{\Lambda} \  d |\spacevec{k}| \ g_{\rm full}(\spacevec k^2)\,\frac{r(|\spacevec k|)}{|\spacevec k|^3}\,, \notag\\
\alpha_0^> & \equiv \frac{3\,\alpha_s\,F_\pi^2\,\bar g_1^{N\!N}}{2\pi\,\Lambda^2}\,, \notag\\
\alpha_2^> &\equiv -\frac{5\,\alpha_s\,F_\pi^2\,\bar g_1^{N\!N}}{8\pi\,\Lambda^4}\,.
\label{eq:def_alphas_C1}
\end{align}

\subsection{Numerical results and discussion}
\label{sec:ALL-subsec:results}
We now turn to the numerical evaluation of the results of the matching expression for $\tilde{\cal C}_1(\mu_\chi,m_s)$ in Eq.~\eqref{eq:mc4} and its low-mass expansion in Eq.~\eqref{eq:taylorexp_C1_NLO}. Following Ref.~\cite{Cirigliano:2021qko}, we use the Kaplan--Steel three-Yukawa potential \cite{Kaplan:1999qa} as our baseline short-range $N\!N$ interaction model, solving numerically the Lippmann--Schwinger (LS) equation to obtain $f_S(\spacevec{q}, \spacevec{p})$, which was then used to evaluate ${\cal I}_C^< (| \spacevec{k}|)$ and $r(|\kk|)$. Fig.~\ref{Fig:C1__vs__ms} shows the behavior of $\tilde{\cal C}_1(\mu_\chi,m_s)$ as a function of $m_s$ for $\mu_\chi=m_\pi$. The figure compares Eq.~\eqref{eq:mc4} (solid blue line) and the low-mass expansion in Eq.~\eqref{eq:taylorexp_C1_NLO} (dashed red line), zooming in the region between $100\MeV$ and $500\MeV$ to appreciate when the two expressions begin to differ.
This corresponds to the sterile neutrino mass at which the approximations in Eq.~\eqref{eq:taylorexp_C1_NLO} are no longer valid, meaning $m_s\gtrsim\lambda$. Notice that the behavior of $\tilde{\cal C}_1(m_s)$ in Eq.~\eqref{eq:mc4} is expected to be corrected by terms of $\Order{(m_s/\Lambda_\chi)}$ that we do not control. Hence, 
a na\"{i}ve uncertainty estimate, $\tilde{\cal C}_1(m_s)\Big(1\pm\Order{(m_s/\Lambda_\chi)}\Big)$, is shown by the blue shaded region in Fig.\ \ref{Fig:C1__vs__ms}.\\

\begin{figure}[!htb]
\centering
\includegraphics[width=0.6\textwidth]{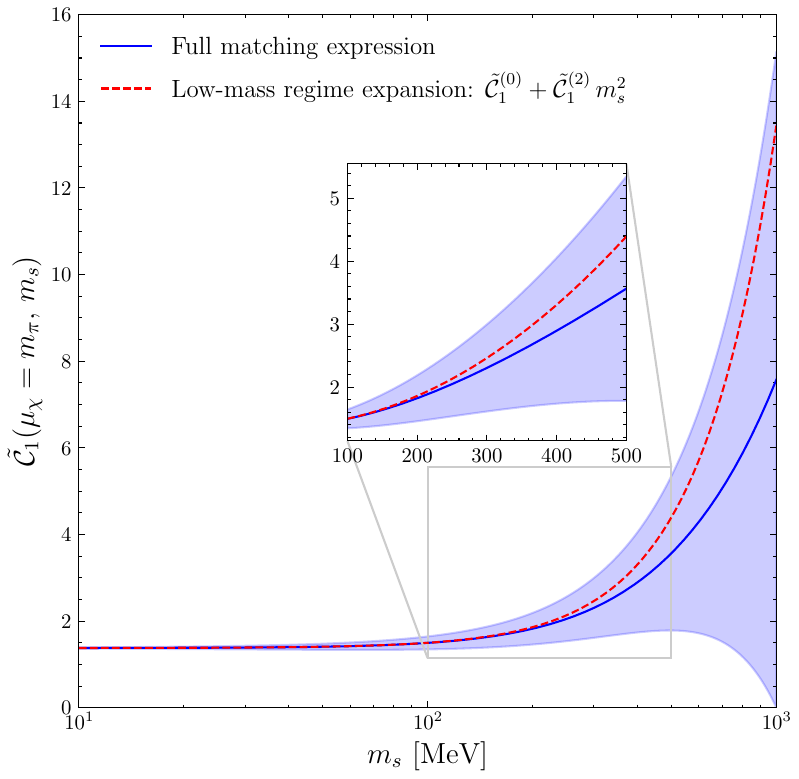}
\caption{$\tilde{\cal C}_1(\mu_\chi,m_s)$ as a function of $m_s$ (in MeV) for $\mu_\chi=m_\pi$. The full matching expression in Eq.~\eqref{eq:mc4} is shown as a solid blue line, whereas the low-mass expansion in Eq.~\eqref{eq:taylorexp_C1_NLO} is shown as a dashed red line. The na\"{i}ve uncertainty band given by $\tilde{\cal C}_1(m_s)\Big(1\pm\Order(m_s/\Lambda_\chi)\Big)$ is shown as a blue shaded region. For the concreteness of this plot, we set $|\spacevec{p}|=|\spacevec{p^\prime}|=1\,{\rm MeV}$, $\Lambda=2\GeV$, $\lambda=100\MeV$, and $\bar g_{1}^{N\!N}=0$.}
\label{Fig:C1__vs__ms}
\end{figure}

For plotting and numerical purposes, the results depicted in Fig.~\ref{Fig:C1__vs__ms} were obtained by setting $|\spacevec{p}|=|\spacevec{p'}|=1\MeV$, $\Lambda=2\GeV$, $\lambda=100\MeV$, and $\bar g_1^{N\!N}=0$. Nonetheless, we have verified that varying $\Lambda\in[1\GeV,\,4\GeV]$ and $\bar g_1^{N\!N}\in[-10,+10]$ does not affect the behavior of $\tilde{\cal C}_1(\mu_\chi,m_s)$ significantly\,\footnote{For $m_s=100\MeV$, the individual variations of $\Lambda$ or $\bar g_{1}^{N\!N}$ will modify the LEC $\sim\pm0.5\%$ with respect to the result in Fig.~\ref{Fig:C1__vs__ms}.}.\\

To ensure the robustness of our results, we have also checked the independence of $\tilde{\cal C}_1(\mu_\chi,m_s)$ on the scale $\lambda$ and the external momenta $\pext$. Fig.~\ref{Fig:C1__vs__lambda} depicts the stability of the LEC under variations of $\lambda$ for three different values of $m_s$; $1\MeV$ (blue lines), $100\MeV$ (red lines), and $3m_\pi$ (green lines). The solid and dashed lines represent the use of Eq.~\eqref{eq:mc4} and its low-mass expansion in Eq.~\eqref{eq:taylorexp_C1_NLO} to compute $\tilde{\cal C}_1(\mu_\chi,m_s)$, respectively. Notice that, even for $m_s=3m_\pi$ (mass value for which the approximated expression differs from the full calculation by $\sim20\%$), the value of $\tilde{\cal C}_1(\mu_\chi,m_s)$ is hardly affected by the choice of the scale $\lambda$. To explore the sensitivity to the external momenta, we solved the LS equation for a different external configuration ($|\spacevec{p}|=1\MeV$, $|\spacevec{p^\prime}|=38\MeV$) to obtain a second kinematic point. As Footnote \ref{fnote:p-dep_r} points out, the numerical calculation of $r(|\kk|)$ requires a fixed $\pext$, leading to a residual momentum dependence in our matching prescription. However, the difference between these two kinematic points corresponds to only a $\sim1\%$ variation in the LEC within the IR region, indicating that the $\pext$ dependence of $\tilde{\cal C}_1$ is negligible.\\

\begin{figure}[!htb]
\centering
\includegraphics[width=0.6\textwidth]{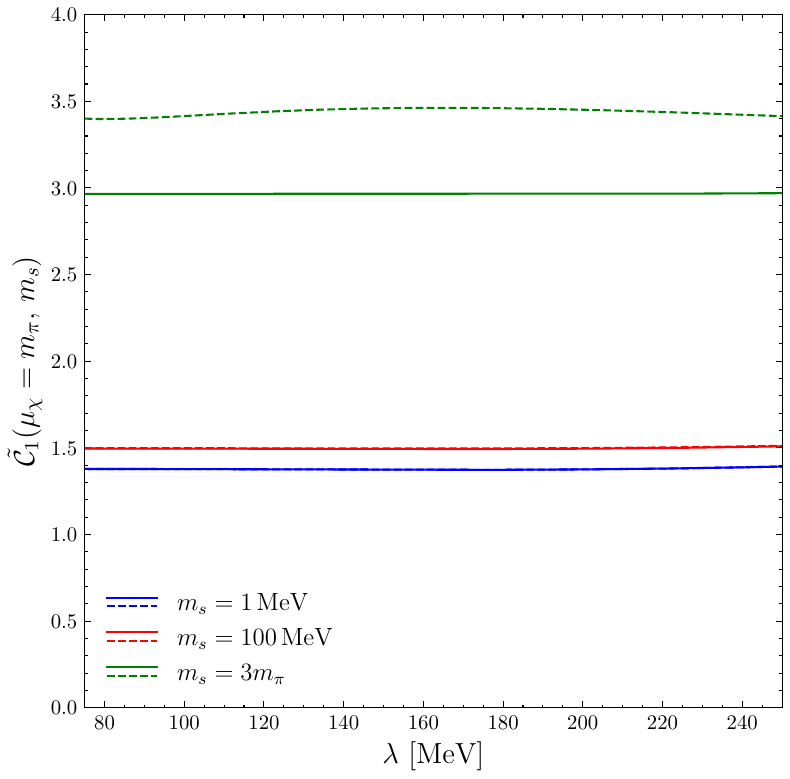}
\caption{$\tilde{\cal C}_1(\mu_\chi,m_s)$ as a function of the scale $\lambda$ (in MeV) for $\mu_\chi=m_\pi$ and three different values of $m_s=\{1\MeV,\,100\MeV,\,3m_\pi\}$. The solid and dashed lines represent the use of the full matching expression in Eq.~\eqref{eq:mc4} and its low-mass expansion in Eq.~\eqref{eq:taylorexp_C1_NLO} to compute $\tilde{\cal C}_1$, respectively. For the concreteness of this plot, we set $|\spacevec{p}|=|\spacevec{p^\prime}|=1\,{\rm MeV}$, $\Lambda=2\GeV$, and $\bar g_1^{N\!N}=0$ . 
}
\label{Fig:C1__vs__lambda}
\end{figure}

Finally, since the results in Eqs.~\eqref{eq:mc4} and \eqref{eq:taylorexp_C1_NLO} agree reasonably well, and using the fact that the LEC has a small sensitivity to the external scales $\Lambda$, $\lambda$, and $\pext$, we can use Eq.~\eqref{eq:taylorexp_C1_NLO} to provide a polynomial expression for $\tilde{\cal C}_1$,

\begin{align}
\tilde{\cal C}_1(\mu_\chi=m_\pi,m_s) \simeq 1.377 + \pfrac{12.062}{{\rm GeV}^2}m_s^2 + \pfrac{-16.735}{{\rm GeV}^4}m_s^4\,,
\label{eq:LEC_C1}
\end{align}
calculating the result for $\Lambda=2\GeV$, $|{\bf p}|=|{\bf p'}|=1\MeV$, $\lambda=100\MeV$, and $\bar g_1^{N\!N}=0$.

\section{Non-minimal scenario: Left-Right amplitude}
\label{sec:AVV_ms}
The Lagrangian in Eq.\ \eqref{eq:Seff0} summarizes all the $\nu$SM interactions relevant  for \NLDBD. This minimal model can be straightforwardly extended to BSM scenarios in which the sterile neutrinos are still the only light new fields, $m_s\lesssim m_W$, but which allow for additional heavy BSM particles, $m_X\gg m_W$. These models can be captured by extending the set of operators in the Lagrangian beyond dimension four, leading to an effective field theory with $\nu_R$ and the SM fields as explicit degrees of freedom \cite{delAguila:2008ir,Cirigliano:2013xha,Liao:2016qyd}. This EFT is the so-called $\nu$SMEFT, which is able to describe a wide array of scenarios, ranging from left-right models to extensions involving leptoquarks and GUTs \cite{Pati:1974yy,Mohapatra:1974hk,Senjanovic:1975rk,Perez:2013osa,Dorsner:2016wpm,Grinstein:2006cg}. \\

In the context of \NLDBD, the most important effect is that the  charged-current weak interactions in Eq.\ \eqref{eq:Seff0} can appear with more general Lorentz structures,
\be
{\cal L}_{\nu{\rm SMEFT}}  \supset  
\sqrt{2} G_F C^{(6)}_{\al\bt} \    (\bar{u} \Gamma_\al d) \, (\bar{e} \Gamma_\bt \nu_{e} )
 +  {\rm h.c.}  ~, 
\label{eq:nuSMEFT}
\ee
where $\Gamma_{\al,\bt}$ denote Dirac structures and the ($\nu$)SM coupling is given by  $C_{\rm VLL} =-2 V_{ud}$, corresponding to $\Gamma_{\rm VL}\otimes \Gamma_{\rm VL} = \gamma_\mu P_L\otimes \gamma^\mu P_L$. Here, we will not dive into all possible $\nu$SMEFT interactions in full generality. Instead, we will restrict the discussion to a minimal extension of the $\nu$SM interactions in which the weak current not only couples to the left-handed quarks but also interacts with right-handed quarks, \emph{i.e.,}\ $C_{\rm VRL}\neq 0$, corresponding to the Dirac structure $\Gamma_{\rm VR}\otimes \Gamma_{\rm VL} = \gamma_\mu P_R\otimes \gamma^\mu P_L$. These interactions can play an important role in models such as left-right symmetric scenarios \cite{Li:2024djp,deVries:2022nyh,Prezeau:2003xn,Li:2020flq}.\\

In this setup, owing to the presence of a coupling involving the right-handed quark current, 
the low-energy $N\!N$ effective Lagrangian to leading order in $Q/\Lambda_\chi$  involves two operators 
\begin{align}
\mathcal L_{\Delta L =2}^{N\!N} &= 
 2 \,  G_F^2 V_{ud}^2  
\, 
\left(    \sum_{i=1}^{3+n} \, U_{ei}^2  \, m_i    \,
{\cal C}_1 (m_i)  \, O_1 \ + \
 \sum_{i=1}^{3+n} \, U_{ei}^2  \, m_i    \, 
{\cal C}_2 (m_i) \,   O_2    
   \right) \,   \bar{e}_L e_L^c   
\nonumber \\
O_1  &= \bar N \mathcal Q^+_L N \, \bar N \mathcal Q^+_L N  
+ \{ L \leftrightarrow R \} \, ,
\nonumber \\
O_2 &=  2 \left( \bar N \mathcal Q^+_L N \bar N \mathcal  Q^+_R N  
-\frac{\textrm{Tr}[\mathcal Q^+_L \mathcal Q^+_R] }{6} \bar N \boldtau N \cdot 
\bar N \boldtau N \right) \, ,
\label{eq:CLR1}
\end{align}
where  
\begin{equation}
\mathcal Q_L^a  = u^\dagger t^a u\,, 
\qquad 
\mathcal Q_R^a  = u t^a u^\dagger\,,
\end{equation}
with $t^+ = (\tau^1 + \i \tau^2)/2$  and  $u  = \exp\big(\i \boldtau \cdot \boldsymbol{\pi}/(2 F_\pi)\big)$ incorporates the pion fields.\\

In the pion sector, we also have a  non-derivative $\Delta L=2$  local operators arising from the $LR$ current correlator:
\begin{align}
\mathcal L_{\Delta L =2}^{\pi \pi }  &=  \left( 8  G_F^2 V_{ud}^2   \  \bar{e}_L e_L^c   \right) \ 
\sum_{i=1}^{3+n}U_{ei}^2\, m_i\,Z(m_i) F_\pi^4 \ 
 \textrm{Tr} [\mathcal Q^{+}_L \mathcal Q^{+}_R ]\,, 
\label{eq:ZLag}
\end{align}
where, at LO in chiral perturbation theory ($\chi$PT), the low-energy constant (LEC) $Z (m_s)$ is related to the electromagnetic 
 pion-mass splitting by 
\begin{equation}
Z(0) e^2 F_\pi^2 = \frac{1}{2}\delta m^2_\pi  = \frac{1}{2} \left( m_{\pi^\pm}^2 - m_{\pi^0}^2 \right).
\label{eq:pion-mass-splitting}
\end{equation}

The goal of this section is to generalize the analysis  of  Ref.~\cite{Cirigliano:2021qko} 
to determine  the $LR$ coupling, ${\cal C}_2 (m_s)$  at $m_s \neq 0$.  
Following  Ref.~\cite{Cirigliano:2021qko},  instead of extracting ${\cal C}_2$ directly, 
we will focus on the vector combination $\mathcal{C}_1 + \mathcal{C}_2$.  
In this case the matching calculation is most easily done by considering the $nn \to pp$ transition in the unphysical theory in which 
the $W$ boson couples to both the left-handed quark current $\bar{u}_L \gamma_\mu d_L$ (as in the Standard Model) and the right-handed current $\bar{u}_R \gamma_\mu d_R$ with equal couplings. As in Sec. \ref{sec:ALL_ms}, we will present the most relevant aspects of the calculation, highlighting the main differences compared with the results obtained in Ref.~\cite{Cirigliano:2021qko} for $m_s=0$.\\

The starting point of the analysis is the  effective action induced by two insertions of the unphysical vector-like weak interaction,
\begin{align}
 S_{{\rm eff}, VV}^{\Delta L = 2} 
=& 
\frac{ 8  G_F^2 V_{ud}^2  
}{2!} 
\sum_{i= 1}^{ 3+n} U_{ei}^2\, m_i
\int d^4 x\,  d^4y   \ 
\bar{e}_L   (x)  \gamma^\mu  \gamma^\nu e_L^c (y)  
\nonumber \\
&\times 
 \int \frac{d^4k}{(2 \pi)^4}   \frac{e^{-\i k \cdot (x - y)}}{k^2 - m_i^2 + \i \epsilon}
\  T \Big\{ 
J_{\mu L}(x)   \,  J_{\nu L}(y)   + 
J_{\mu R}(x)   \, J_{\nu R}(y)   + 
2  J_{\mu L}(x) \,  J_{\nu R}(y)   
\Big\}\,,
\label{eq:SVVeff1-v0}
\end{align}
with $J_{\mu L}  = \bar u_L \gamma_\mu d_L$ and $J_{\mu R}  = \bar u_R \gamma_\mu d_R$.
To estimate the relevant LECs, we will equate full theory and EFT matrix elements of the above effective action  between 
hadronic states $\langle h_f|$ and $| h_i \rangle$: 
\begin{align}
\langle e_1^- e_2^-  h_f |   S_{\rm eff, V\!V}^{\Delta L = 2}  | h_i   \rangle 
&=  (2 \pi)^4 \, \delta^{(4)}  (p_f - p_i)  
\Big( 4  G_F^2 V_{ud}^2 
\   \bar{u}_L   (p_1)  u_L^c (p_2)  \Big)  
\times 
\sum_{i=1}^{3+n} U_{ei}^2 \, m_i  \, 
 {\cal A}_{VV}^{fi} (m_i)\,,\notag   \\
{\cal A}_{VV}^{fi}  (m_i)  & =
2\,  g^{\mu \nu}  \,
 \int \frac{d^4k}{(2 \pi)^4} \, 
\frac{ T^{fi}_{\mu \nu} (k,p)
}{ k^2 - m_i^2 + \i \epsilon}\, . 
\nonumber \\
T^{fi}_{\mu \nu}  (k,p)  
  &=   \int d^4 r  \  
 e^{\i k \cdot r} \, 
 \langle h_f (p) |  T  \Big\{  J_\mu^V(+r/2)   \ J_\nu^V(- r/2)  \Big\}    | h_i (p) \rangle ~, 
\end{align}
where $J_\nu^V \equiv J_\mu^L + J_\mu^R = \bar u \gamma_\mu d$. 
In what follows, we first study  the simpler $\pi^- \to \pi^+$ transition, which determines $Z(m_s)$ and is required as input to the 
matching calculation of $nn\to pp$, which determines ${\cal C}_1 (m_s) + {\cal C}_2 (m_s)$.

\subsection[$\pi^- \to \pi^+$ vector-like amplitude]{$\boldsymbol{\pi^- \to \pi^+}$ vector-like amplitude}
\label{sec:AVV_ms-subsec:Z}

For the simpler $\pi^- \to \pi^+$ transition,  we perform the matching at zero external momentum. 
The full QCD amplitude is given by
\begin{align}
{\cal A}_{VV}^{\pi \pi}  (m_s)  \Big \vert_{\rm QCD}  = 
2 \, \int \frac{d^dk}{(2 \pi)^d} \,  \frac{ g^{\mu \nu} T_{\mu \nu}^{\pi \pi} (k,0)}{k^2 - m_s^2 + \i  \epsilon} \,.
\end{align}

This amplitude, as a function of $m_s$, has been determined using Lattice QCD \cite{Tuo:2022hft}. Here, however, we model the amplitude by using the parametrization that was introduced in Ref.~\cite{Cirigliano:2021qko} for the hadronic tensor $T_{\mu \nu}^{\pi \pi} (k,0)$. 
At low  and intermediate momenta, we  include the pion pole contribution with  appropriate insertions of the pion  vector form factor, 
assumed to be dominated by a single resonance (the $\rho$ meson) with mass $M_V$. 
At high momentum, we adopt the  form of the  hadronic tensor dictated by the QCD operator product expansion, 
which depends on the  $\pi^- \to \pi^+$ matrix element of a local operator, denoted by $g_{LR}^{\pi \pi}$~\cite{Cirigliano:2021qko}.   
After splitting  up the integration over $d|\spacevec{k}|$ into low- and high-momentum regions, we arrive at the representation 
\begin{eqnarray}
{\cal A}_{VV}^{\pi \pi}  (m_s)  \Big \vert_{\rm QCD}  &=& {\cal A}_{\pi \pi}^<(m_s^2) + {\cal A}_{\pi \pi}^>(m_s^2)
\,,\notag\\
&&\notag\\
{\cal A}_{\pi \pi} ^<(m_s^2) &=&  \frac{3}{\pi^2}\pfrac{M_V^2}{M_V^2 - m_s^2}^2\, \int_0^\Lambda  \, d |\spacevec{k}|  \, \spacevec{k}^2\ 
\frac{(\omega_V - \omegak)^2(2\omega_V + \omegak)}{\omegak \omega_V^3} \,,\notag
\\
{\cal A}_{\pi \pi}^>(m_s^2) &=& (16 F_\pi^2)\,\frac{\alpha_s (\mu)\, g_{LR}^{\pi \pi} (\mu)}{4\pi} \, \int_{\Lambda}^{\infty}   d |\spacevec{k}|  \ \pfrac{2|\spacevec k| + \omegak}{\omegak |\spacevec k|(|\spacevec k| + \omegak)^2}\,, 
\label{eq:Zmatch}
\end{eqnarray}
where  $\omega_V = \sqrt{\kk^2 + M_V^2}$ and  $\mu$ is the QCD renormalization scale.
We note that in the chiral limit, the contribution of the pion intermediate state in the hadronic tensor leads to  non-analytic behavior in $m_s$,  namely 
\begin{align}
{\cal A}_{\pi \pi} ^<(m_s^2) 
&\simeq A_{\pi \pi}^< (0) + \frac{3m_s^2}{2\pi^2} \ln \pfrac{m_s^2}{4\Lambda^2}+{\cal O}(m_s^2)\,.
\label{eq:pipi1}
\end{align}

The EFT side of the matching condition also displays this non-analytic behavior, which arises from loops, in such 
a way that the LEC $Z(m_s)$  has a polynomial behavior in $m_s^2$\,\footnote{More precisely, the quark level Lagrangian should induce Eq.\ \eqref{eq:ZLag} with $Z(0)$, along with terms $\sim m_s^n$  each of which should come with its own LEC, $Z^{(n)}$. This is equivalent to treating the combination of these terms as an `effective LEC' with $m_s$ dependence, $Z(m_s^2)$.}. 
In fact, in order to capture the $\Order(m_s^2)$ terms,  one-loop contributions need to be included in the $\chi$PT amplitude. 
Generalizing  the results of Ref.~\cite{Cirigliano:2017tvr}, we find the following  EFT amplitude 
\begin{eqnarray}
 {\cal A}_{VV}^{\pi \pi}  (m_s)  \Big \vert_{\chi \rm PT}  &=&   
16F_\pi^2  \left[ Z(m_s^2)+
\frac{m_s^2}{32\pi^2F_\pi^2}\left(3\ln\pfrac{m_s^2}{\mu_\chi^2}-1\right) \right] \,. 
\label{eq:pipi2}
\end{eqnarray}

The matching relation
\be
 {\cal A}_{\rm V\!V}^{\pi \pi}  (m_s)  \Big \vert_{\chi \rm PT}  \ = \  {\cal A}_{\rm V\!V}^{\pi \pi}  (m_s)  \Big \vert_{\rm QCD}  
\ee
leads to 
\be
Z(m_s^2) = \frac{1}{16F_\pi^2} 
\, \bigg( 
{\cal A}_{\pi \pi}^<(m_s^2) + {\cal A}_{\pi \pi}^>(m_s^2)
\bigg) 
+ \frac{m_s^2}{32\pi^2F_\pi^2}\left[1 - 3\ln\pfrac{m_s^2}{\mu_\chi^2} \right]  \,,
\label{eq:ZmatchF}
\ee
which thanks to Eqs.~\eqref{eq:pipi1} and \eqref{eq:pipi2}  is free of 
non-analytic terms 
in  $m_s$ as $m_s \to 0$.\\

The above analysis confirms that the chiral EFT reproduces the same non-analytic $m_s$ dependence that we find on the full-theory side, as expected from general principles. These results are useful when determining $Z(m_s)$ from a full-theory representation. In addition, a similar kind of cancellation of non-analytic terms is expected to occur in the $nn\to pp$ matching that we discuss in the next section.

\subsection[$nn \to pp$ vector-like amplitude]{$\boldsymbol{nn \to pp}$ vector-like amplitude}
\label{sec:AVV_ms-subsec:nn2pp}

Denoting by ${\cal A}_{LL,RR,LR}$ the $nn \to pp$ amplitudes generated by  neutrino exchange between two left-handed, 
two right-handed, and  a left-handed and a right-handed  current, respectively,  
the $nn \to pp$ vector-like amplitude can be written as  
\begin{equation}
{\cal A}_{VV} = 2 {\cal A}_{LL} + 2 {\cal A}_{LR}\ ~, 
\label{eq:MV}
\end{equation}
where we made use of parity invariance of the strong interactions,  which implies ${\cal A}_{RR} = {\cal A}_{LL}$. 
Note that ${\cal A}_{LL} \equiv {\cal A}_\nu$, {\it i.e.,} the amplitude studied in Section~\ref{sec:ALL_ms}.

\subsubsection{Chiral EFT amplitude}
\label{sec:AVV_ms-subsec:nn2pp-subsubsec:chiEFT}
At LO in chiral EFT,  the $N\!N \to pp$ vector amplitude takes the form
\begin{equation}
{\cal A}_{VV}^{\chi {\rm EFT}} = 2 {\cal A}_{LL}^{N\!N} +  2 {\cal A}_{LR}^{N\!N} +  {\cal A}_{VV}^{\pi \pi}\,, 
\label{eq:MVEFT}
\end{equation}
where ${\cal A}_{LL}^{N\!N}$ is proportional to $\mathcal A_\nu^{\rm \chi EFT}$ in Eq.~\eqref{eq:AchiPT}. ${\cal A}^{N\!N}_{LR}$  can be obtained  from ${\cal A}^{N\!N}_{LL}$ by flipping the sign of the  ``$A\times A$'' axial contribution (since $L\times L\sim V\times V+A\times A$, while $L\times R\sim V\times V-A\times A$), which in practice amounts to setting  $g_A^2 \to - g_A^2$ everywhere and replacing ${\cal C}_1 \to {\cal C}_2$ in the counterterm amplitude. The amplitude ${\cal A}_{VV}^{\pi \pi}$ has the same form of $ {\cal A}_\nu^{\chi {\rm EFT}}$, except for the fact that in each diagram, the neutrino propagator is replaced by a pion propagator with one insertion of $Z(m_s)$ that converts a $\pi^-$ into a $\pi^+$ (see Fig.\ 9 in Ref.~\cite{Cirigliano:2021qko}). 
This amplitude does  not involve explicit neutrinos so that the $m_s$ dependence of the potential comes purely from the LEC $Z(m_s^2)$. 

In complete analogy to the  $LL$ case, the matching condition for the $VV$ amplitude  involves 
only the real part of the singular component of the amplitude ${\cal A}_C$ (see Fig.~\ref{fig:diagrams} for the topological definition of ${\cal A}_C$). Additionally,  certain parts of the amplitude are needed beyond LO in order to capture the right IR behavior, as discussed for the $LL$ case in 
Section~\ref{sec:matching}.
All in all, we find 
\begin{eqnarray}
\mathcal A_C^{\rm sing}(\mu_\chi,m_s)\Big|_{VV} 
&=& -2\pfrac{m_N}{4\pi}^2\Big[A(m_s,\mu_\chi,\pext)+2g_A^2\,Z(m_s^2)\,A(0,\mu_\chi,\pext)\Big]\,,
\label{eq:ACsingVV}
\end{eqnarray}
with the function
\begin{eqnarray}
A(m_s,\mu_\chi,\pext) &\equiv& \log\pfrac{\mu_\chi^2}{m_s^2+\pext^2} + \dNLO\,\pi\,m_s + 1\,.
\label{eq:def_A}
\end{eqnarray}

The UV divergence in $\mathcal A_C^{\rm sing}(\mu_\chi,m_s)|_{VV}$ is removed by the combination of counterterms given by 
$4/C^2 \, \Big({\cal C}_1 (m_s^2) + {\cal C}_2 (m_s^2)\Big)$. 

\subsubsection{Full amplitude}
As in Sec. \ref{sec:ALL_ms}, for matching purposes, we only need the singular parts of the amplitude ${\cal A}_C^<$, which can be decomposed 
as in   Eq.~\eqref{eq:MVEFT},
\begin{equation}
{\cal A}_C^{<,{\rm sing}}(m_s) \Big \vert_{VV}
= 2 \left(  {\cal A}_C^{<,{\rm sing}}(m_s) \Big \vert_{LL}^{N\!N}
+ {\cal A}_C^{<,{\rm sing}}(m_s) \Big \vert_{LR}^{N\!N}
\right) + 
{\cal A}_C^{<, {\rm sing}}(m_s)\Big|_{VV}^{\pi\pi}\,.
\label{eq:ACVVm}
\end{equation}

As in the case of the $LL$ amplitudes, some care must be taken to ensure that non-analytic terms in external momenta and $m_s$ cancel in the 
matching condition.  We next discuss the contributions from the two-nucleon intermediate state and from the pion exchange term. 
\\

\noindent {\bf Two-nucleon contribution:} ${\cal A}_C^{<,{\rm sing}} \vert_{LL}^{N\!N}$ is obtained by combining the results in Eqs.~\eqref{eq:integrals}, and \eqref{eq:aLLm}--\eqref{eq:ACmsing}, while  ${\cal A}_C^{<,{\rm sing}} \vert_{LR}^{N\!N}$ is obtained  simply by flipping the sign of the 
axial terms in ${\cal A}_C^{<,{\rm sing}} \vert_{LL}^{N\!N}$. 
Introducing the auxiliary scale $\lambda$ such that  $|\spacevec p|_{\rm ext} \ll \lambda\lesssim m_\pi$, and breaking up the momentum integration 
in two pieces as in  Section~\ref{sect:fullLL} we arrive at 
\begin{equation}
2\left(  {\cal A}_C^{<,{\rm sing}}(m_s) \Big \vert_{LL}^{N\!N}
+ {\cal A}_C^{<,{\rm sing}}(m_s) \Big \vert_{LR}^{N\!N} \right) 
= 2\ \frac{m_N^2}{(4 \pi)^2} \, \left[  \int_0^{\lambda}+ \int_{\lambda}^\Lambda \right] \,  d |\spacevec{k}|  \   a^{N\!N}_< ( | \spacevec{k} |,m_s) \,, 
\label{eq:ACNNm}
\end{equation}
with
\begin{align}
\int_0^{\lambda} d |\spacevec{k}| \ a^{N\!N}_< ( | \spacevec{k} |,m_s) &= 
-g_{\rm full}^{VV}(0)\Bigg[-\dNLO\,\lambda + \dNLO\,m_s\tan^{-1}\pfrac{\lambda}{m_s} 
+\ \frac{1}{2}\log\pfrac{m_s^2+\lambda^2}{m_s^2+\pext^2}\Bigg]\,, 
\nonumber \\
\int_{\lambda}^{\Lambda} d |\spacevec{k}| \ a^{N\!N}_< ( | \spacevec{k} |,m_s) &= -\int_{\lambda}^{\Lambda} d |\spacevec{k}| \ g_{\rm full}^{VV}(\spacevec k^2)\pfrac{|\spacevec k|\,r(|\spacevec k|)}{\spacevec k^2+m_s^2}\,, 
\end{align}
and 
\begin{equation}
g_{\rm full}^{VV}(\kk^2) \equiv 
2 g_V^2  (\spacevec{k}^2) +   \frac{\spacevec{k}^2 g_M^2 (\spacevec{k}^2)}{m_N^2}\,.
\end{equation}

The high momentum contribution is analogous to the $m_s=0$ case, and we have omitted a detailed discussion here. 
\\

\noindent {\bf Pion exchange contribution:} 
In the full theory, the equivalent of the EFT terms proportional to $Z(m_s)$  (called $ {\cal A}_{VV}^{\pi \pi} $) is generated by the insertion of the two vector currents and the pion exchanged between the two nucleons (see Fig.~10 in Ref.~\cite{Cirigliano:2021qko}). 
For the low- and intermediate-momentum region, this results in 
\begin{eqnarray}
{\cal A}_C^{<, {\rm sing}}(m_s)\Big|^{\pi\pi}_{VV}  = 
- m_N^2 \, \frac{g_A^2}{F_\pi^2}
\int \frac{d^4k}{(2 \pi)^4} \ \frac{1}{k^2 - m_s^2 + \i\epsilon}  \,  
\int \frac{d^3 \spacevec{q}}{(2 \pi)^3}  \
\frac{1}{\spacevec{q}^2}
\,  {\cal I}_C^< ( |\spacevec{q}|) 
\, g^{\mu \nu} T_{\mu \nu}^{\pi \pi} (k,p)  ~. 
\label{eq:ACpipim}
\end{eqnarray}

The strategy to compute the integral in Eq.~\eqref{eq:ACpipim} is described in detail in Sec. 6.3.2 of 
Ref.~\cite{Cirigliano:2021qko}. In short, it builds up the general result from the simpler case of $F_\pi^V(k^2) =1$ and $ {\cal I}_C^< ( |\spacevec{q}|) = {\cal I}_C(|\spacevec{q}|)$.
In the chiral limit ($m_\pi\to0$), we have 
\begin{align}
{\cal A}_C^{<, {\rm sing}}(m_s)\Big|^{\pi\pi}_{VV} =& 2\,\frac{m_N^2}{(4 \pi)^2}\Bigg[-4g_A^2\, \frac{{\cal A}^<_{\pi \pi}(m_s^2)}{(16F_\pi^2)}\,\log\pfrac{\nu_\chi}{\pext} + \int_0^\Lambda d |\spacevec{k}|  \ \Bigg(  a_<^{\pi\pi}(|\spacevec k|,m_s,\nu_\chi)\notag\\
&+\ \delta a_<^{\pi\pi}(|\spacevec k|,m_s) + \Delta a_<^{\pi\pi}(|\spacevec k|,m_s) \Bigg)  \Bigg]\,,\label{eq:AVVpipi<_presubtraction}\\
a_<^{\pi\pi}(|\spacevec k|,m_s,\nu_\chi) =& -\frac{8g_A^2}{(4\pi F_\pi)^2}\pfrac{\omega_V^2-\kk^2}{\omega_V^2-\omegak^2}^2\,|\spacevec k|\Bigg[\log\pfrac{\omegak+|\spacevec k|}{\omegak} - \log\pfrac{\omega_V+|\spacevec k|}{\omega_V}\notag\\
&+\ \frac{|\spacevec k|}{\omegak}\Bigg\{- \pfrac{\omega_V-\omegak}{\omega_V} + 3\log\pfrac{\omegak+|\spacevec k|}{\nu_\chi} 
-\ 3\pfrac{\omegak}{\omega_V}\log\pfrac{\omega_V+|\spacevec k|}{\nu_\chi} 
\notag\\
&- \frac{\omegak(\omega_V^2-\omegak^2)}{2\omega_V^3}\Bigg(3\log\pfrac{\omega_V+|\spacevec k|}{\nu_\chi}
-\ \frac{ 3 \omega_V+  |\spacevec k|}{\omega_V+|\spacevec k|}\Bigg)\Bigg\}\Bigg]\,,
\label{eq:apipi2}
\\
\delta a_<^{\pi \pi} (|\spacevec{k}|,m_s)  =&  8 g_A^2\frac{ | \spacevec{k}|}{(4 \pi F_\pi)^2}  
\ \int_{\lambda}^\Lambda d |\spacevec{q}| \, \pfrac{1 - r (|\spacevec{q}|)}{|\spacevec{q}|} \, g (|\spacevec{k}|, |\spacevec{q}|) \,, \label{eq:dapipi}\\
\Delta a_<^{\pi \pi} (|\spacevec{k}|,m_s) 
=&  8 g_A^2\frac{\dNLO\, | \spacevec{k}|}{(4 \pi F_\pi)^2}  
\ \int_0^{\lambda} d |\spacevec{q}| \ g (|\spacevec{k}|, |\spacevec{q}|) \,. \label{eq:Dapipi} 
\end{align}

The details of the function $g(|\spacevec{k}|,|\spacevec{q}|)$ are given in Appendix \ref{app:ACpipim_details}.
A few comments are in order:
\begin{itemize}
\item With the aim to explicitly reconstruct in the full amplitude the term in the EFT amplitude proportional to $Z(m_s^2) \times \log (\mu_\chi^2/ \pext^2)$  (see Eq.~\eqref{eq:ACsingVV}), we have isolated  the logarithmic term in $\pext$, 
proportional to  ${\cal A}_{\pi \pi}^< (m_s^2)$ (see Eq.~\eqref{eq:Zmatch}). 
In the process, we have introduced the arbitrary scale $\nu_\chi$. 

\item Up to the above manipulation, the term $a_<^{\pi\pi}$  accounts for the effects of the pion form factors, whereas the terms $\delta a_<^{\pi\pi}$ and $\Delta a_<^{\pi\pi}$ come from the inclusion of the HOS form factor effects by rewriting ${\cal I}_C^< ( |\spacevec{q}|)$ in Eq.~\eqref{eq:ACpipim} using the procedure discussed in Sec. \ref{sec:matching}.  The related function $g(|\kk|,|\spacevec{q}|)$ is given explicitly  in Eq.~\eqref{eq:AppA_g_deff}. 

\item 
Similarly to the case of   ${\cal A}_{\pi \pi}^<(m_s^2)$  discussed in Section~\ref{sec:AVV_ms-subsec:Z}, 
 the low-momentum   amplitude  ${\cal A}_C^{<,{\rm sing}}\vert^{\pi\pi}_{VV}$ in Eq.~\eqref{eq:AVVpipi<_presubtraction} 
 contains non-analytic terms in $m_s$ and $\pext$. 
 Such terms arise from both the  ${\cal A}_{\pi \pi}^<(m_s^2)$   and  
$ a_<^{\pi\pi}(|\spacevec k|,m_s,\nu_\chi)$ terms in Eq.~\eqref{eq:AVVpipi<_presubtraction}.  
The same non-analytic terms appear in the EFT amplitude and are eventually canceled in the matching condition for the low-energy couplings, 
as shown explicitly in the simpler example of $Z(m_s^2)$ in Section~\ref{sec:AVV_ms-subsec:Z}. 
For the $nn \to pp$ case, some of the relevant  EFT  diagrams for the $LL$ case at $m_s=0$ were calculated in Ref.~\cite{Cirigliano:2017tvr}, and similar calculations can be performed for the $LR$ case with nonzero $m_s$ as well. 
However,  the full N2LO calculation of the $nn\to pp$ amplitude is beyond the scope of this work 
and we, therefore, take a more pragmatic approach: we simply assume the cancellation takes place and  subtract any non-analytic terms in $m_s$ from 
Eq~ \eqref{eq:apipi2} in the matching condition. 

\item  We depart from the analysis of Ref.~\cite{Cirigliano:2021qko}  in the following detail: in the evaluation of the RHS of Eq.~\eqref{eq:ACpipim} 
we do not include the  evanescent term of  $g^{\mu \nu} T_{\mu \nu}^{\pi \pi} (k,p)$, which is proportional to $d-4$.  
Upon integration, this would generate extra finite pieces that cancel between Eqs.~\eqref{eq:apipi2} and \eqref{eq:dapipi}. 
Ref.~\cite{Cirigliano:2021qko} kept only the finite piece in  Eq.~\eqref{eq:apipi2}, missing 
the cancellation. 
This implies that our result for $\tilde{\cal C}_1 + \tilde{\cal C}_2$ at $m_s=0$ is shifted by $\sim-0.6$ compared to  the corresponding result in Ref.~\cite{Cirigliano:2021qko}.

\end{itemize}

To estimate the contribution from the high-momentum region,   ${\cal A}_{VV}^>$, 
we use the OPE results
from Ref.~\cite{Cirigliano:2021qko} to obtain the following expression for ${\cal A}_C^{>,{\rm sing}}\vert_{VV}$:\\
\begin{align}
{\cal A}_C^{>,{\rm sing}}(\tilde\nu_\chi,m_s) \Big \vert_{VV} &= 2\,\frac{m_N^2}{(4\pi)^2}\left[-4 g_A^2 \, \frac{{\cal A}_{\pi \pi}^>(m_s^2)}{(16F_\pi)^2}\, \log \pfrac{\tilde\nu_\chi}{\pext} + \int_\Lambda^\infty d|\spacevec k|\ a^{VV}_>(|\spacevec k|,m_s,\tilde\nu_\chi)\right]\,,\notag\\
a^{VV}_>(|\spacevec k|,m_s,\tilde\nu_\chi) &= \frac{2 \alpha_s}{ \pi} \,(4\pi F_\pi)^2\, \pfrac{2|\spacevec{k}| + \omegak}{\omegak|\spacevec{k}|(|\spacevec{k}| + \omegak)^2}  \
\left(2\bar g_{LR}^{N\!N} (\tilde\nu_\chi) - \frac{\bar g_{LR}^{\pi \pi} \,  g_A^2 }{4}\right)\,,\label{eq:ACVVp}
\end{align}
where $\tilde\nu_\chi$ is the chiral EFT renormalization scale used in evaluating 
the matrix element of the local operator that controls the high-momentum tail of the amplitude (for simplicity, we take $\tilde\nu_\chi=\nu_\chi$; see Ref.~\cite{Cirigliano:2021qko} for details) and we
 have  isolated the IR-dependent contribution proportional to  ${\cal A}_{\pi \pi}^> (m_s^2) \equiv 16F_\pi^2\,Z^>(m_s^2)$   (see  Eq.~\eqref{eq:Zmatch}). 
 Using the lattice QCD results of Ref.~\cite{Nicholson:2018mwc},
 we find $\bar g_{LR}^{\pi \pi}  = 8.23$ in the $\overline{\rm MS}$ scheme at $\mu = 2\GeV$. While both $\bar g_{LR}^{\pi \pi}$ and $\bar{g}_{LR}^{N\!N} $ are expected to be $\Order(1)$, given the large numerical value of $\bar g_{LR}^{\pi \pi} $, for the nucleon coupling we will assume $\bar{g}_{LR}^{N\!N} \in [-10,+10]$.

\subsection[Matching condition for ${{\cal C}_1(m_s) + {\cal C}_2(m_s)}$]{Matching condition for $\boldsymbol{{\cal C}_1(m_s) + {\cal C}_2(m_s)}$}
\label{sec:AVV_ms-subsec:matching}

Rescaling the  vector-like couplings and amplitudes in the same manner as in Eqs.~\eqref{eq:match3} and \eqref{eq:match4}, and using the matching condition for $Z(m_s^2) $ in Eq.~\eqref{eq:ZmatchF},
one arrives at 
\begin{align}
2\,\Big(\tilde C_1 + \tilde C_2\Big)(\mu_\chi,
m_s) =& \Big[A(m_s,\mu_\chi,\pext) + 2g_A^2\,Z(m_s^2)\,A(0,\mu_\chi,\pext)\Big]\notag\\
&- 4 g_A^2 \, Z(m_s^2)   \, \log \pfrac{\nu_\chi}{\pext} \notag\\
&+ \int_0^{\Lambda}   d |\spacevec{k}| \ a^{VV}_< ( | \spacevec{k} |, m_s, \nu_\chi)  + \int_{\Lambda}^\infty     d |\spacevec{k}| \  a^{VV}_> ( | \spacevec{k} |, m_s, \nu_\chi)\notag\\
&- \sigma_1^{\pi\pi}\,m_s^2\log\pfrac{m_s}{\Lambda_S} - \sigma_2^{\pi\pi}\,m_s^2\log\pfrac{m_s}{\Lambda_S}^2\,, \label{eq:match4V}
\end{align}
with  $A(m_s,\mu_\chi,\pext)$ given in Eq.~\eqref{eq:def_A},  
\begin{align}
a^{VV}_< ( | \spacevec{k} |, m_s, \nu_\chi) &\equiv 
  a^{N\!N}_< ( | \spacevec{k} |, m_s)   + a^{\pi \pi}_< ( | \spacevec{k} |, m_s,\nu_\chi)   + \delta a^{\pi \pi}_< ( | \spacevec{k} |,m_s) 
   + \Delta a^{\pi \pi}_< ( | \spacevec{k} |,m_s)\,, 
\label{eq:aVVm2}
\end{align}
and the coefficients 
\begin{align}
\sigma_1^{\pi\pi} &= \frac{2g_A^2}{(4\pi F_\pi)^2}\left[1+6\log\pfrac{\nu_\chi}{\Lambda_S} \right]\,,\notag\\
\sigma_2^{\pi\pi} &= -\frac{6g_A^2}{(4\pi F_\pi)^2}\,,
\end{align}
determined by the  $m_s\ll\Lambda$ expansion of the following integral  
\begin{align}
\label{eq:a_<_pipi_expansion}
\int_0^{\Lambda}   d |\spacevec{k}| \ a^{\pi\pi}_< ( | \spacevec{k} |, m_s,\nu_\chi) 
&= \alpha_0^{<,\pi\pi} + \alpha_2^{<,\pi\pi}\,m_s^2 + \sigma_1^{\pi\pi}\,m_s^2\log\pfrac{m_s}{\Lambda_S} + \sigma_2^{\pi\pi}\,m_s^2\log\pfrac{m_s}{\Lambda_S}^2\, . 
\end{align}

Eq.~\eqref{eq:a_<_pipi_expansion}  implements our pragmatic approach to the cancellation of infrared effects in the matching condition, 
see discussion in the third bullet below Eq.~\eqref{eq:Dapipi}. 
Although $\Lambda_S$  is an arbitrary scale, the fact that the corresponding chiral EFT loop diagrams can only involve terms $\sim\log(m_s/m_\pi)$ or $\sim\log(m_s/\nu_\chi)$, we will vary the value of $\Lambda_S$ around $m_\pi$ to explore the sensitivity of the LECs to this procedure. \\

Combining all the previous results, we see how the linear piece in $m_s$ cancels out, and the LECs behave as expected,
\begin{align}
\Big(\tilde C_1 + \tilde C_2\Big) (\mu_\chi,
m_s) &= \tilde{\cal C}_{1+2}^{(0)} + \tilde{\cal C}_{1+2}^{(2)}\,m_s^2 + \Order\left(m_s^4
\right)\,,
\label{eq:match5V}
\end{align}
with   expansion coefficients are given by
\begin{align}
\tilde{\cal C}_{1+2}^{(0)} 
=& \frac{1}{2}\Bigg[1 + 2\dNLO\,\lambda + 2\log\pfrac{\mu_\chi}{\lambda} - \int_{\lambda}^{\Lambda}d|\spacevec k|\ g_{\rm full}^{VV}(\spacevec k^2)\,\frac{r(|\spacevec k|)}{|\spacevec k|} + 2g_A^2\left[1 + 2\log\pfrac{\mu_\chi}{\nu_\chi}\right]Z(0)\Bigg]\notag\\
&+\ \alpha_0^{<,\pi\pi} + \delta\alpha_0^{<,\pi\pi} + \Delta\alpha_0^{<,\pi\pi} + \alpha_0^{>,VV}\Bigg] \,,\notag\\
\tilde{\cal C}_{1+2}^{(2)} 
=& \frac{1}{2}\Bigg[\frac{2\dNLO}{\lambda} - \frac{1}{\lambda^2} + \int_{\lambda}^{\Lambda}d|\spacevec k|\ g_{\rm full}^{VV}(\spacevec k^2)\,\frac{r(|\spacevec k|)}{|\spacevec k|^3} + 2g_A^2\left[1 + 2\log\pfrac{\mu_\chi}{\nu_\chi}\right]Z'(0)\notag\\
&+\ \alpha_2^{<,\pi\pi} + \delta\alpha_2^{<,\pi\pi} + \Delta\alpha_2^{<,\pi\pi} + \alpha_2^{>,VV}\Bigg]\,.
\label{eq:match5V_terms}
\end{align}

In evaluating  $\tilde{\cal C}_1 + \tilde{\cal C}_2$ from the matching condition in Eq.~\eqref{eq:aVVm2} or from its expanded form 
of Eq.~\eqref{eq:match5V_terms} one needs as input the low-energy constant $Z(m_s^2)$, whose value at $m_s=0$ is related to the pion electromagnetic 
mass splitting.  To minimize the model dependence, we take $Z(m_s=0)$ from  experiment and only use  Eq.~\eqref{eq:ZmatchF}
to determine the $m_s^2$-dependent terms in $Z(m_s^2)$. Using the model description of Eq.~\eqref{eq:ZmatchF} to evaluate $Z(0)$
would decrease  $\tilde{\cal C}_{1+2}^{(0)} $ by  $\approx 0.3$.\\

\subsection{Numerical results and discussion}

We now present numerical results corresponding to the matching results in Eq.~\eqref{eq:match4V} and the low-mass analysis detailed in Eq.~\eqref{eq:match5V}. In addition to the calculation considerations described in Sec. \ref{sec:ALL-subsec:results}, we vary the subtraction scale $\Lambda_S\in[m_\pi/2,\,2m_\pi]$ to explore the sensitivity of the LECs to this energy scale introduced by our procedure. Although we present our results for a particular choice of $|\spacevec{p}|=|\spacevec{p'}|=1\MeV$, $\Lambda=2\GeV$, $\lambda=100\MeV$, $\nu_\chi=m_\pi$, and $\bar g_{LR}^{N\!N}=0$, we have verified that varying $\Lambda\in[1\GeV,\,4\GeV]$ and $\bar g_{LR}^{N\!N}\in[-10,10]$ do not affect our results significantly.\\

\begin{figure}[!htb]
\centering
\includegraphics[width=0.6\textwidth]{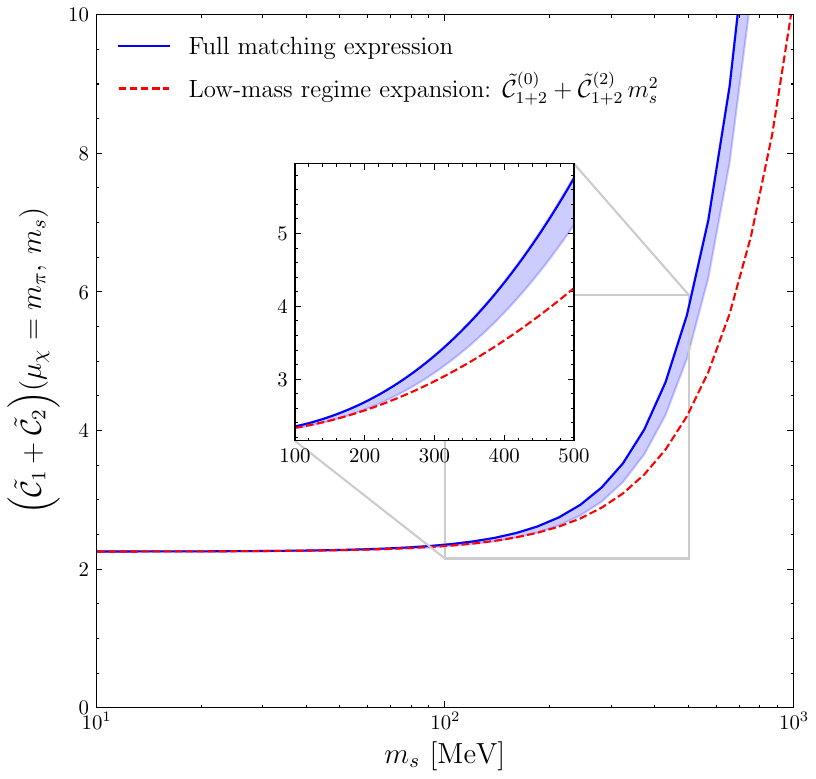}
\caption{$\Big(\tilde{\cal C}_1+\tilde{\cal C}_2\Big)(\mu_\chi,m_s)$ as a function of $m_s$ (in MeV) for $\mu_\chi=m_\pi$. The blue band corresponds to the effects of varying $\Lambda_S\in[m_\pi/2,\,2m_\pi]$. The full matching expression in Eq.~\eqref{eq:match4V} is shown as a solid blue line, whereas the low-mass regime expansion in Eq.~\eqref{eq:match5V} is shown as a dashed red line. For the concreteness of this plot, we set $|\spacevec{p}|=|\spacevec{p^\prime}|=1\,{\rm MeV}$, $\Lambda=2\GeV$, $\lambda=100\MeV$, $\nu_\chi=\Lambda_S=m_\pi$, and $\bar g_{LR}^{N\!N}=0$.
}
\label{Fig:C1C2__vs__ms}
\end{figure}

In Fig.~\ref{Fig:C1C2__vs__ms} we present the behavior of $(\tilde{\cal C}_1+\tilde{\cal C}_2)(\mu_\chi,m_s)$ as a function of $m_s$ for $\mu_\chi=m_\pi$. The figure compares the full matching expression in Eq.~\eqref{eq:match4V} (solid blue line) and the low-mass expansion in Eq.~\eqref{eq:match5V} (dashed red line), also depicting the effects of the subtraction scale $\Lambda_S$ in the LECs. The blue band covers the interval $\Lambda_S\in[m_\pi/2,\,2m_\pi]$, whereas the corresponding solid line presents the representative value $\Lambda_S=m_\pi$. 
Note that the value of $(\tilde{\cal C}_1+\tilde{\cal C}_2)(\mu_\chi,m_s=0)$ is smaller than the one in Ref.~\cite{Cirigliano:2021qko}, 
as discussed below Eq.~\eqref{eq:Dapipi}.

Similarly to the $LL$ case in Sec. \ref{sec:ALL-subsec:results}, since the results in Eqs.~\eqref{eq:match4V} and \eqref{eq:match5V} agree reasonably well, we can use Eq.~\eqref{eq:match5V_terms} to provide a polynomial expression for $\tilde{\cal C}_1+\tilde{\cal C}_2$,
\begin{align}
\Big(\tilde{\cal C}_1+\tilde{\cal C}_2\Big)(\mu_\chi=m_\pi,m_s) \simeq 2.253 + \pfrac{7.993}{{\rm GeV}^2}m_s^2\,,
\label{eq:LEC_C1C2}
\end{align}
calculating the result for $\Lambda=2\GeV$, $|{\bf p}|=|{\bf p'}|=1\MeV$, $\lambda=100\MeV$, $\nu_\chi=\Lambda_S=m_\pi$, and $\bar g_{LR}^{N\!N}=0$. \\

\section{Impact on \boldNLDBD\ half-life}
\label{sec:impact}

In this section, we discuss the impact of the determination of ${\cal C}_1 (m_s)$ on \NLDBD\  decay rates.
To evaluate \NLDBD\ in the presence of light sterile neutrinos, we follow the EFT approach of Ref.~\cite{Dekens:2024hlz} in which the inverse half-life can be written as
\begin{align}
\left(T_{1/2}^{0\nu}\right)^{-1}=g_A^4 G_{01}\Bigg| \sum_{i=1}^{3+n}\frac{U_{ei}^2 \,m_i}{m_e} A_\nu(m_i)\Bigg|^2\,,
\end{align}
where $g_A = 1.2754\pm 0.0013$ \cite{ParticleDataGroup:2024cfk} is the axial-vector charge of the nucleon, $G_{01} = 1.5\cdot 10^{-14}$ yr$^{-1}$ (for $^{136}$Xe) is a phase-space factor \cite{Horoi:2017gmj}, and $A_{\nu}$ is the amplitude. 
The latter receives contributions from (sterile) neutrinos with different momentum scalings, depending on $m_s$,
\begin{align}\label{eq:fullint}
A_\nu (m_i) = \begin{cases}
A_\nu^{\rm (pot,<)}+A_\nu^{\text{(hard)}}(m_i)+A_\nu^{\text{(usoft)}}(m_i) \,,&  m_i <  100 \text{ MeV}\,, \\
A_\nu^{\rm (pot)}+A_\nu^{\text{(hard)}}(m_i)\,,& 100 \text{ MeV} \le m_i <  2 \text{ GeV}\,,\\
A_\nu^{\text{(9)}}(m_i)\,,& {\rm }\,  2 \text{ GeV} \le m_i \,.
\end{cases}
\end{align}
Here $A_\nu^{\rm (9)}$ captures contributions from sterile neutrinos that can be integrated out at the quark level, whenever $m_s\gtrsim 2$ GeV, in which case they can effectively be described by dimension-nine operators. Instead, the superscript $\rm (pot,<)$ and (pot) denote the contributions from neutrinos with potential momenta, $k_0\ll |\spacevec k|\sim m_\pi$. The effects of these neutrinos are determined by Nuclear Matrix Elements, similar to those induced by the long-range exchange of active neutrinos. 
$A_\nu^{\rm (usoft)}$ represents contributions from ultrasoft neutrinos with momenta or the order of the excited-state energies or the $Q$ value of the process, $k_0\sim|\spacevec k|\sim$MeV. These effects depend on transition matrix elements between the initial/final state and intermediate nuclear states, as well as their energies. All of these contributions are discussed in detail in Ref.~\cite{Dekens:2024hlz} and are expected to be dominant when $m_s\lesssim m_\pi$ (potential and ultrasoft) or when $m_s\gtrsim 2$ GeV ($A_\nu^{(9)}$).\\

Finally, $A_\nu^{\rm (hard)}$ captures the effects of hard neutrinos with momenta of $k_0\sim |\spacevec k|\sim 1$ GeV. It is this momentum region that is responsible for the LEC ${\cal C}_1$ and where the results discussed in this paper play a role. The corresponding amplitude can be written as 
\begin{align}\label{eq:hardAmp}
A_\nu^{\rm (hard)}(m_s) = -2\pfrac{m_\pi^2 M_{F,sd}}{g_A^2} {\cal C}_1(m_s)\,,
\end{align}
where $M_{F,sd}$ is the nuclear matrix element of the contact operator. Here we will use the value obtained in the shell-model calculation of Ref.~\cite{Jokiniemi:2021qqv}, $M_{F,sd}=-1.94$ for the decay of $^{136}$Xe.\\

$\chi$EFT predicts that ${\cal C}_1 (m_s)$ admits a series expansion in the neutrino mass for $m_s\ll\Lambda_\chi\simeq 1$ GeV. However, for larger $m_s\sim 1$ GeV this is no longer a good description as the expansion in $m_s/\Lambda_\chi$ breaks down. In fact, one expects the amplitude, and therefore ${\cal C}_1$, to behave as $1/m_s^2$ in the limit of $m_s\gg\Lambda_\chi$. To ensure both the high- and low-mass behavior can be satisfied, Ref.~\cite{Dekens:2024hlz} employed an interpolation formula of the form,
\begin{equation}\label{eq:gnu_int}
{\cal C}_1(m_s) ={\cal C}_1(0) \frac{1+ (m_s/m_c)^2\,{\rm sign}(m_d^2)}{1 + (m_s/m_c)^2(m_s^2/|m_d^2|)}\,.
\end{equation}

Here $m_{c}$ parametrizes the $m_s^2$ dependence in the low-mass region and was set to $m_c=1$ GeV, based on an NDA estimate. Instead, $m_d^2$ is a real (but not necessarily positive) number that determines the large-mass behavior and was fixed by demanding that the full amplitude reproduces $A_\nu^{(9)}$ at $m_s=2$ GeV, which resulted in $m_d^2 \simeq (146\,{\rm  MeV})^2$ for $^{136}$Xe.\\

In principle, one could simply update the value of  $m_c$, using $\tilde {\cal C}_1^{(2)}$ determined above, to gauge the impact of the results derived here. However, the above prescription turns over to the $1/m_s^2$ behavior at rather low values of $m_s\simeq 100-200$ MeV, where we would still expect the expansion of $\tilde {\cal C}_1$ in terms of $m_s$ to hold. We therefore extend the above parametrization of ${\cal C}_1(m_s)$ so that it more closely follows its series expansion up to higher values of $m_s$. In particular, we will use an ansatz of the form, 
\begin{align}\label{eq:newinterpolation}
{\cal C}_1(m_s) ={\cal C}_1(0) \frac{1+ a_2 m_s^2+ a_4 m_s^4}{1 + |b_4| m_s^4+|b_6| m_s^6}\,,
\end{align}
where two of the $a_i$ and $b_i$ coefficients are fixed through the ${\cal C}^{(2,4)}_1$ coefficients determined given in Eq.\ \eqref{eq:LEC_C1}. 
The remaining two coefficients are obtained by demanding that $A_\nu^{(9)}$ is reproduced at $m_s=2$ GeV and that ${\cal C}_1(m_s)$ has a maximum at $m_s\simeq 3 m_\pi$. This last constraint is used as a proxy for the value of $m_s$ up to which the $m_s$ expansion can be trusted, as the large-$m_s$ behavior, ${\cal C}_1\sim 1/m_s^2$, takes over after the maximum is reached.\,\footnote{Explicitly, this fit gives rise to the following values for the coefficients appearing in the interpolation, $a_2 = 8.8\, {\rm GeV}^{-2}$, $a_4 = -1.9\, {\rm GeV}^{-4}$, $b_4 = 10\, {\rm GeV}^{-4}$, and $b_6 = 11\, {\rm GeV}^{-6}$.} \\

\begin{figure}[!t]
\centering
\includegraphics[width=0.6\textwidth]{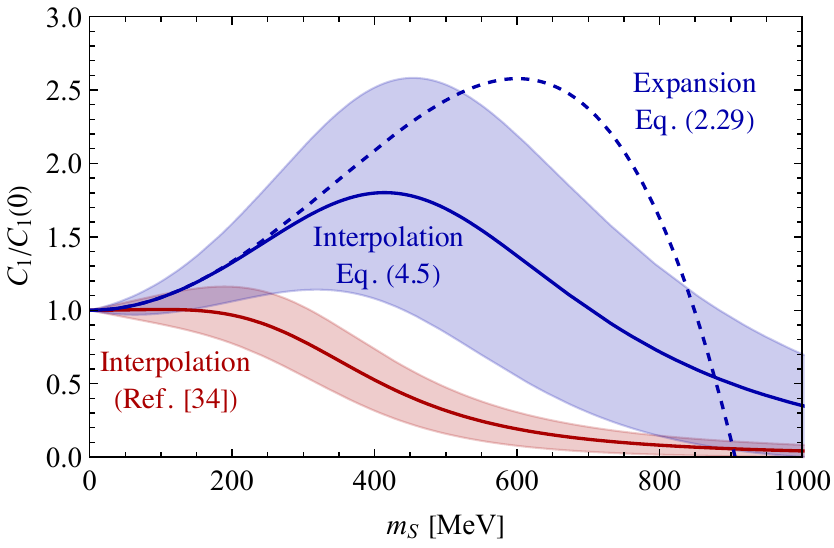}
\caption{$\tilde{\cal C}_1(m_s)/\tilde{\cal C}_1(0)$ as a function of $m_s$ (in MeV), using different approximations and interpolations. The dashed blue line shows the expansion up to $m_s^4$, while the blue solid line depicts the result of Eq.\ \eqref{eq:newinterpolation}, both of which use the results derived in this work. Instead, the red line represents the interpolation of Eq.\ \eqref{eq:gnu_int} that was used in Ref.~\cite{Dekens:2024hlz}.
The shaded regions depict a naive estimate of the uncertainty given by ${\cal C}_1(m_s)(1\pm m_s/\Lambda_\chi)$.
}
\label{Fig:C1interpolation}
\end{figure}

Following the above prescription, we obtain the plot in Fig.\ \ref{Fig:C1interpolation}. Here, the dashed line shows the expansion up to $m_s^4$, the blue line depicts the result of Eq.\ \eqref{eq:newinterpolation}, while the red line shows the result of Ref.~\cite{Dekens:2024hlz} for comparison.
The uncertainties, estimated by ${\cal C}_1(m_s)(1\pm m_s/\Lambda_\chi)$, are shown by the shaded bands. One sees that the updated interpolation does indeed follow the expansion to around $m_s\simeq 3 m_\pi$, after which it turns over. In addition, the results obtained here suggest that ${\cal C}_1$ first increases before turning over, while this effect is hardly visible for the red line. This is due to the fact that NDA suggests a  significantly smaller $\tilde {\cal C}_1^{(2)}$ than the one obtained here.

\subsection{Impact in a simplified scenario}
To illustrate the impact of the determination of ${\cal C}_1$ in phenomenological applications, here we consider a toy model in which the SM is extended with a single $\nu_R$, corresponding to the so-called $3+1$ scenario. This is a toy model as it predicts two massless neutrino states, however, it captures most of the features that are relevant to $0\nu\bt\bt$ of more realistic models without the complication of adding more unknown parameters. To evaluate \NLDBD\ half lives, we will make use of several relations that hold in this specific model, namely, $m_{1,2}=0$ and 
\begin{align}\label{eq:31cancel}
m_3U_{e3}^2 = -m_4U_{e4}^2 = -\frac{1}{3}\frac{m_3 m_4}{m_3 +m_4}e^{\i\phi}\,,
\end{align}
where $\phi$ is a combination of Majorana phases that drops out in the rate. After setting the mass of the only massive active neutrino to $m_3 = \sqrt{\Delta m_{32}^2} \simeq 0.05$ eV, the half life becomes a function of $m_s= m_4$. 
Note that Eq.\ \eqref{eq:31cancel} implies a cancellation between the contributions of $\nu_3$ and $\nu_4$ to \NLDBD. This feature holds more generally than just the $3+1$ scenario, since, as discussed in Section \ref{sec:intro},  all models that extend the SM with $n$ sterile neutrinos predict $m_{\bt\bt} =\sum_{i=1}^{3+n} U_{ei}^2\, m_i = 0$.
\\

With these assumptions, the decay rate can be evaluated as a function of $m_s$, the result of which is shown in the left and right panels of Fig.\ \ref{Fig:C1impact} which focus on different ranges of $m_s$. The blue line shows the results obtained here, using the interpolation of Eq.\ \eqref{eq:newinterpolation}, while the red line depicts the results obtained in Ref.~\cite{Dekens:2024hlz}. As can be seen from the left panel, the half life is barely affected by the change in ${\cal C}_1$ for $m_4\ll m_\pi$, where the scaling of the rate with $m_s$ is roughly described by $\Gamma^{0\nu}_{1/2}\sim m_4^3 U_{e4}^2\sim m_4^2$. As can be seen from the right panel in Fig.\ \ref{Fig:C1impact}, however, the new value of ${\cal C}_1^{(2)}$ obtained here does have a significant impact on the half life, especially for $m_s\gtrsim m_\pi$. The shaded regions show the uncertainty due to ${\cal C}_1$, which we estimate by ${\cal C}_1(m_s)(1\pm m_s/\Lambda_\chi)$. We stress that these bands do not account for other sources of uncertainty, such as those in the nuclear matrix elements, see Ref.\ \cite{Dekens:2024hlz} for a more detailed discussion. The fact that the blue uncertainty band shrinks to a point near $m_s\simeq 700$ MeV is due to the fact that ${\cal C}_1(700\, {\rm MeV})-{\cal C}_1(0)\simeq 0$, as can also be seen from Fig.\ \ref{Fig:C1interpolation}.
Although the uncertainty on our results grows with increasing $m_s$, it is nevertheless clear that the larger value of ${\cal C}_1^{(2)}$ leads to considerably longer half lives, which differ from the red line by up to a factor of  $6$.\\

The reason for the decrease in the amplitude (increase in the half life) is that we find a larger value of ${\cal C}_1^{(2)}$ than the one that was assumed in Ref.~\cite{Dekens:2024hlz}, which was based on NDA. Although the long- and short-distance contributions add up in the scenario with $3$ active neutrinos with Majorana masses (induced by the dimension-five Weinberg operators), they interfere destructively in the minimal extension of the SM, as a result of the relation $\sum_{i=1}^{3+n} U_{ei}^2 \, m_i = 0$. The larger value of ${\cal C}_1^{(2)}$ leads to a more effective cancellation between the long- and short-distance contributions and thereby a larger half-life.\\

More realistic models that can describe the experimental masses and mixing angles of the active neutrinos, such as the $3+2$ case discussed in Ref.\ \cite{deVries:2024rfh}
or $3+3$ scenarios, will be affected by ${\cal C}_1^{(2)}$ in a similar way. In these cases, the short-distance contribution from each added neutrino is still described by Eq.\ \eqref{eq:hardAmp}. As a result, the impact of ${\cal C}_1^{(2)}$ for each $\nu_i$ will be analogous to the features depicted in Fig.\ \ref{Fig:C1impact} and most significant in the range $\Lambda_\chi\gtrsim m_i\gtrsim m_\pi$.

\begin{figure}[!t]
\centering
\includegraphics[width=0.49\textwidth]{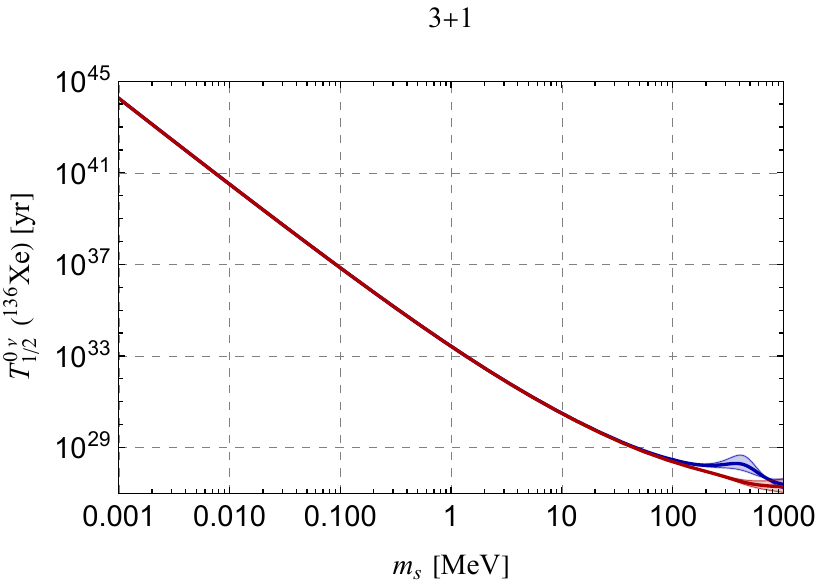}
\includegraphics[width=0.49\textwidth]{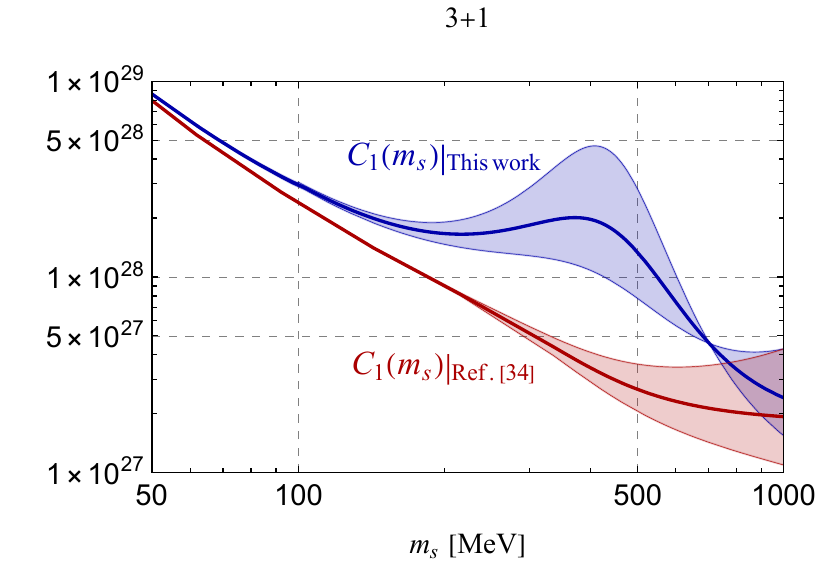}
\caption{
The half-life of \NLDBD\ in the $3+1$ scenario as a function of the sterile mass $m_4=m_s$, with the left and right panels showing different regions ranges of $m_s$. Here, the blue line employs the interpolation of Eq.\ \eqref{eq:newinterpolation} and the determination of $\tilde {\cal C}_1 (m_s)$ derived in this work. The red line depicts the results of Ref.~\cite{Dekens:2024hlz}, which employed the simpler interpolation of Eq.\ \eqref{eq:gnu_int} and an NDA estimate for $\tilde {\cal C}_1^{(2)}$. The shaded regions depict the error induced by the uncertainty in ${\cal C}_1$, which we estimate by ${\cal C}_1(m_s)(1\pm m_s/\Lambda_\chi)$.
}
\label{Fig:C1impact}
\end{figure}

\section{Conclusions and outlook} 
\label{sec:conclusion}

Neutrinoless double-beta decay probes a wide range of lepton-number-violating interactions and mechanisms for generating neutrino masses.
A simple and well-motivated class of models of Majorana neutrino mass involves the introduction of 
$n$ right-handed Majorana neutrinos, which are singlets under the Standard Model gauge group. 
The phenomenological implications of this model depend on the mass scale of these sterile states. 
In this work we have studied the implications of 
sterile neutrinos with masses $m_s$ below the GeV scale on neutrinoless double beta decay. 
The present work fills a gap in our knowledge by determining 
the $m_s$ dependence of the short-range $nn \to pp$ effective couplings that appear to leading order 
in the chiral EFT description of neutrinoless double beta decay with sterile neutrinos~\cite{Dekens:2020ttz,Dekens:2023iyc,Dekens:2024hlz}. 
Our main results and their impact can be summarized as follows: 
\begin{itemize}

\item Generalizing  the Cottingham-like matching strategy developed in Refs.~\cite{Cirigliano:2020dmx,Cirigliano:2021qko}, 
we have determined the $nn \to pp$ effective couplings appearing in Eqs.~\eqref{eq:C1v1} and \eqref{eq:CLR1}, 
relevant for the minimal $\nu$SM scenario and left-right symmetric extensions of the Standard Model, respectively. 

\item The main subtleties in the matching analysis arise from the infrared behavior, \emph{i.e.,}\ when $\pext\to0$ or $m_s \to 0$. 
In the discussion, we have highlighted the key steps needed to 
make sure that  the effective couplings do not depend on $\pext$ and are polynomials in $m_s^2$, 
in accordance with the general principles of effective field theory. 

\item The resulting effective couplings have, by construction, the correct anomalous dimension 
and display mild dependence on the arbitrary matching scales introduced in intermediate steps in the analysis. 
{
Similarly to the $m_s \to 0$ case~\cite{Cirigliano:2020dmx, Cirigliano:2021qko}, 
substantial sources of uncertainty in ${\cal C}_{1,2} (m_s)$ arise from missing 
inelastic $N\!N\pi$ intermediate states 
and model-dependence of the short-range $N\!N$ interaction, resulting in an overall 30-50\% fractional error.  
The recent analysis of Ref.~\cite{VanGoffrier:2024lmo} argues that 
the inclusion of the inelastic channel $N\!N\pi$ significantly reduces the uncertainty 
while shifting the central value by less than 10\%. This gives us confidence that 
our current result, in the limit $m_s \to 0$,  is reliable at the level of 20-30\%. 
The most important 
departure from the $m_s=0$ analysis}
is due to the fact that the effective theory description involves an expansion in $m_s/\Lambda_\chi$. This implies our uncertainties from missing higher-order terms increase with increasing $m_s$ and our results can only be applied below the breakdown scale of $m_s\sim \Lambda_\chi$.
For ease of implementation in phenomenological studies, we have provided simple polynomial expansions for $\tilde {\cal C}_{1,2} (m_s)$  in Eqs.~\eqref{eq:LEC_C1}  and \eqref{eq:LEC_C1C2}.

\item  We have illustrated the impact of our new results in the $\nu$SM, showing 
that the $m_s$-dependence of the couplings differs significantly from naive 
dimensional estimates~\cite{Dekens:2023iyc,Dekens:2024hlz}. 
For $m_s$ is in the 200-800 MeV range, our results imply a significantly longer half-life (almost 
one order of magnitude), keeping all other parameters fixed. This will affect   
the interpretation of a positive or null $0\nu \beta \beta$ experimental signal in terms of sterile neutrinos. 

\end{itemize}

In this work, we have focused on the regime $m_s < \Order({\rm GeV})$. 
For $m_s > \Order({\rm GeV})$, the heavy neutrinos should be integrated out after which their effect 
is captured by contact interactions, with couplings scaling $\propto 1/m_s^2$, 
and overall coefficients determined by non-perturbative physics. 
The methods used here and in Refs.~\cite{Cirigliano:2020dmx,Cirigliano:2021qko} can be extended  
to study the $m_s > \Order({\rm GeV})$ regime as well. 
In addition, given the recent progress in $\pi^- \to \pi^+$ matrix elements~\cite{Nicholson:2018mwc,Detmold:2022jwu,Tuo:2022hft,Tuo:2019bue,Feng:2018pdq} 
and in the formalism for the two nucleon system~\cite{Davoudi:2020gxs}, 
we expect that  lattice QCD studies will provide insight on the $nn \to pp$ effective couplings 
for a broad range of $m_s$.

\bigskip

\noindent {\bf Acknowledgments}--- 
We thank Jordy de Vries, Martin Hoferichter, and Emanuele Mereghetti for valuable discussions. 
This work was supported by the INT's U.S. Department of Energy grant No. DE-FG02-00ER41132.\\

\appendix

\section{Chiral EFT amplitude to NLO: $\boldsymbol{m_s}$ dependence} 
\label{sec:appNLO}

The low-momentum contribution to the full amplitude  in Eq.~\eqref{eq:IC_lowmomentum} 
leads to a term proportional to  $m_s \times \dNLO$, which
 is not captured by the leading order  EFT expression~\eqref{eq:chiEFT}. 
A linear dependence on $m_s$  arises on the EFT side at NLO, as described below. \\

To  NLO, the  $nn \to pp e^-e^-$ amplitude in chiral EFT takes the form~\cite{Cirigliano:2019vdj} 
\begin{align}
\mathcal A^{\rm NLO}_{\nu} 
=& \mathcal A_A 
+ \chi^+_{\spacevec p^\prime}(\spacevec 0) \left( K_{E^\prime} + K^{(1)}_{E^\prime}\right)
 \mathcal A_B         
+ \mathcal {\bar A}_B \left(K_E + K_E^{(1)}\right) \chi^+_{\spacevec p}(\spacevec 0) 
\nonumber \\ 
&+\ 
\chi^+_{\spacevec p^\prime}(\spacevec 0) \left(   K_{E^\prime} + K^{(1)}_{E^\prime} \right)  
\left(\mathcal A_C + \frac{2{g^{N\!N}_\nu}}{
C^2} \right) 
\left(K_E + K_E^{(1)} \right) \chi^+_{\spacevec p}(\spacevec 0) 
\nonumber \\
&+\ \chi^+_{\spacevec p^\prime}(\spacevec 0) \, K_{E^\prime} \, \mathcal A^{(1)}_B
+ \mathcal {\bar A}^{(1)}_B  \, K_E \, \chi^+_{\spacevec p}(\spacevec 0)
+ \chi^+_{\spacevec p^\prime}(\spacevec 0) \, K_{E^\prime} \, \mathcal A^{(1)}_C 
\, K_E \, \chi^+_{\spacevec p}(\spacevec 0)
\, .  
\label{eq:amplitudeNLO}
\end{align}

For details and definitions of all quantities, we point the reader to Ref.~\cite{Cirigliano:2019vdj}.
Here, the superscript (1) indicates the NLO pieces.
Here, We focus only on the terms from the ${\cal A}_C$ topology relevant to our discussion (we are not attempting 
a full matching to NLO). 
The $\chi$EFT version of $\mathcal A_C^{(1)}$ is given by
\begin{align}
\mathcal A_C^{(1)} =&  
\left(-\frac{4g_{\nu}^{N\!N}}{C^2}\frac{C_2}{C}+\frac{g_{2\nu}^{N\!N}}{C^2}\right)
\pfrac{\spacevec p^2 + \spacevec p^{\prime\,2}}{2} 
-2m_NV_\pi(0)\pfrac{g_{2\nu}^{N\!N}}{C^2}+\pfrac{C_2}{C^2}m_NV_{\nu L}(0) 
\,, 
\label{AC(1)chi}
\end{align}
where $V_{\nu L}(0)$ and $V_{\pi}(0)$ are the pion and neutrino potentials at $r=0$~\cite{Cirigliano:2019vdj}. 
Assuming that the counterterm contributions only give subleading $m_s$ dependence compared to that of $g_\nu^{N\!N}$ in the LO $\mathcal A_C$ amplitude, the most significant $m_s$ dependence comes from the last term. The integral is given by
\begin{align}
V_{\nu L}(0)=&\int\frac{d^3\spacevec k}{(2\pi)^3}\ \frac{1}{\spacevec k^2+m_s^2}\left[1+2g_A^2+g_A^2\frac{m_\pi^2}{(\spacevec k^2+m_\pi^2)^2}\right]\,.
\end{align}

In the matching procedure, we only kept the most divergent parts of $\mathcal A_C$ and dropped more convergent contributions from pion exchange in the LNV and strong potentials. Applying the same logic here, we conclude that only the term $\sim\!(1+2g_A^2)$ contributes, giving the following result in the power divergence subtraction (PDS)
scheme \cite{Kaplan:1998we, Kaplan:1998tg}
\begin{align}
V^{\rm sing}_{\nu L}(0) &= \pfrac{1+2g_A^2}{4\pi}(\mu_\chi-m_s)\,.
\end{align}

This result allows us to identify the $m_s$-dependent piece from the NLO contribution, leading to
\begin{align}
\operatorname{Re}\left(\bar {\cal A}_{C}^{\rm (1), sing}(\mu_\chi, m_s)-\bar{\cal A}^{\rm (1), sing}_{C}(\mu_\chi, 0)\right) =& -(1+2g_A^2) \pfrac{4\pi}{m_N}\pfrac{ C_2}{C^2}m_s\notag\\
=&-\frac{(1+2g_A^2)}{2}\,\dNLO\,\pi\,m_s\,,
\end{align}
which matches the linear term in $m_s$ in the full theory, Eq.~\eqref{eq:IC_lowmomentum}, after plugging in the $\chi$EFT result for $\dNLO$ of Eq.~\eqref{eq:dNLO}.

\section{Function $\boldsymbol{g(|\spacevec{k}|,|\spacevec{q}|)}$}
\label{app:ACpipim_details}
The function $g (|\spacevec{k}|, |\spacevec{q}|) $ is defined as follows,
\begin{align}
g(|\spacevec{k}|, |\spacevec{q}|)&=\begin{cases}
g_- (|\spacevec{k}|, |\spacevec{q}|)\ ,\qquad \text{if} \quad |\spacevec{q}|<|\spacevec{k}|\\
g_+ (|\spacevec{k}|, |\spacevec{q}|)\ ,\qquad \text{if} \quad |\spacevec{k}|<|\spacevec{q}|
\end{cases},
\label{eq:AppA_g_deff}
\end{align}
where the functions $g_\pm(|\spacevec k|,|\spacevec q|)$ satisfy the relation
\begin{align}
g_\pm(|\spacevec k|,|\spacevec q|) =& \pfrac{\omega_V^2-\spacevec k^2}{\omega_V^2-\omegak^2}^2\Bigg[\mathcal{G}_\pm(\omegak;|\spacevec k|,|\spacevec q|)-\mathcal{G}_\pm(\omega_V;|\spacevec k|,|\spacevec q|)\notag\\
&+\frac{(\omega_V^2-\omegak^2)}{2\omega_V}\dfrac{\partial \mathcal{G}_\pm}{\partial \omega}(\omega_V;|\spacevec k|,|\spacevec q|)\Bigg]\,.
\label{eq:AppA_gplusminus}
\end{align}

Here, we have defined $\omega_X\equiv\sqrt{\kk^2+m_X^2}$ and
\begin{align}
\mathcal{G}_-(\omega;|\spacevec k|,|\spacevec q|) =& \frac{1}{2\omega {|\spacevec q|}}\left[2|\spacevec q|(2|\spacevec k|+\omega) + (\omega^2-{\bf k}^2+2{\bf q}^2)\log\left(1-\frac{2|\spacevec q|}{|\spacevec k| + |\spacevec q| + \omega}\right)\right]\,,\notag\\
\mathcal{G}_+(\omega;|\spacevec k|,|\spacevec q|) =& \frac{1}{2\omega {|\spacevec q|}}\left[2|\spacevec k|(2|\spacevec q|+\omega) + (\omega^2-{\bf k}^2+2{\bf q}^2)\log\left(1-\frac{2|\spacevec k|}{|\spacevec k| + |\spacevec q| + \omega}\right)\right]\,.
\end{align}

Finally, for the low-mass expansion of the $\delta a^{\pi\pi},\Delta a^{\pi\pi}$ terms in Eq.~\eqref{eq:match5V_terms}, it is useful to express these functions as a series in $m_s$:
\begin{align}
g_{\pm}(|\spacevec k|,|\spacevec q|) = g_{\pm}^{(0)}(|\spacevec k|,|\spacevec q|) + g_{\pm}^{(2)}(|\spacevec k|,|\spacevec q|)\,m_s^2 + \Order(m_s^4)\,,\label{eq:AppA_gplusminus_expansion}
\end{align}
with
\begin{align}
g_{\pm}^{(0)}(|\spacevec k|,|\spacevec q|) =& \mathcal{G}_{\pm}(|\spacevec k|;|\spacevec k|,|\spacevec q|) - \mathcal{G}_{\pm}(\omega_V;|\spacevec k|,|\spacevec q|) + \frac{(\omega_V^2-\spacevec k^2)}{2\omega_V}\frac{\partial \mathcal G_{\pm}}{\partial\omega}(\omega_V;|\spacevec k|,|\spacevec q|)\,,\notag\\
g_{\pm}^{(2)}(|\spacevec k|,|\spacevec q|) =& \frac{1}{2|\spacevec k|\omega_V(\spacevec k^2-\omega_V^2)} \bigg[|\spacevec k|(\spacevec k^2-\omega_V^2)\frac{\partial \mathcal G_{\pm}}{\partial\omega}(\omega_V;|\spacevec k|,|\spacevec q|) \notag\\
&+ \omega_V(\spacevec k^2-\omega_V^2)\frac{\partial \mathcal G_{\pm}}{\partial\omega}(|\spacevec k|;|\spacevec k|,|\spacevec q|) \notag\\
&- 4|\spacevec k|\omega_V\Big(\mathcal{G}_{\pm}(|\spacevec k|;|\spacevec k|,|\spacevec q|) + \mathcal{G}_{\pm}(\omega_V;|\spacevec k|,|\spacevec q|)\Big)\bigg]\,.
\end{align}

\bibliographystyle{JHEP}
\bibliography{draft_v2}

\end{document}